\documentclass[fleqn,usenatbib]{mnras}

\usepackage{newtxtext,newtxmath}

\usepackage[T1]{fontenc}

\DeclareRobustCommand{\VAN}[3]{#2}
\let\VANthebibliography\thebibliography
\def\thebibliography{\DeclareRobustCommand{\VAN}[3]{##3}\VANthebibliography}

\usepackage{caption}
\usepackage{subcaption}

\usepackage{graphicx}	
\usepackage{amsmath}	
\usepackage{multirow}   
\usepackage{longtable}
\usepackage{pdflscape}
\usepackage{xcolor}

\defcitealias{drake2017}{D17}

\title[LAEs and CGM enrichment]{Connecting CGM enrichment with Lyman $\alpha$ emitters at $2.9<z<6.7$}

\author[A. M. Sebastian et al.]{
Alma Maria Sebastian,$^{1}$\thanks{E-mail: asebastian@swin.edu.au}
Emma Ryan-Weber,$^{1}$
Rebecca L. Davies,$^{1}$
Romain A. Meyer$^{2}$ and
\newauthor Valentina D'Odorico$^{3,4}$
\\
$^{1}$Centre for Astrophysics and Supercomputing, Swinburne University of Technology, John Street, Hawthorn, Victoria, 3122, Australia\\
$^{2}$Department of Astronomy,University of Geneva, Chemin Pegasi 51, 1290 Versoix, Switzerland
\\
$^{3}$INAF-Osservatorio Astronomico di Trieste, Via Tiepolo 11, I-34143 Trieste, Italy\\
$^{4}$IFPU-Institute for Fundamental Physics of the Universe, via Beirut 2, I-34151 Trieste, Italy
}
\date{Accepted XXX. Received YYY; in original form ZZZ}

\pubyear{\the\year{}}

\begin{document}
\label{firstpage}
\pagerange{\pageref{firstpage}--\pageref{lastpage}}
\maketitle

\begin{abstract}
We present the results of a blind search for Lyman $\alpha$ emitters (LAEs) in three deep archival $z\sim6$ quasar fields from VLT/MUSE using state-of-the-art detection algorithms. We explore their connection with absorbers-particularly \ion{C}{iv} and \ion{Mg}{ii}-in the circumgalactic medium (CGM) from the E-XQR-30 survey. We detect 156 LAEs at $2.9<z<6.7$ with luminosities ranging from log $L[\text{ergs}^{-1}]=$ 41.3 to 43.2. We find 34 and 14 galaxy associations with \ion{C}{iv} and \ion{Mg}{ii} absorption respectively at $3.4<z<5.8$ within a line of sight velocity window of $\pm1000~\text{km}^{-1}$ and impact parameter of $<250$ pkpc. These systems have a weak anti-correlation with respect to the absorber strength-impact parameter relation. No \ion{Mg}{ii} systems are found within the virial radii of any LAE while four \ion{C}{iv} absorbers are located within the virial radii of an LAE suggesting that low ionisation gas has a lower covering fraction. The LAEs have mild overdensity ratios of 1.7 and 1.9 around \ion{C}{iv} and \ion{Mg}{ii} respectively. The stellar mass upper limits of $\text{log}~M_*<10.7~\text{M}_\odot$ estimated using Keck/NIRC2 imaging indicate that a low-mass, faint population of galaxies pollutes the CGM with metals. This paper serves as a pilot analysis for the forthcoming REQUIEM survey, an ESO Large Program on high redshift deep quasar fields. 
\end{abstract}




\section{Introduction}
\label{sec:intro}

The circumgalactic medium (CGM) acts as a reservoir of fuel for star formation in galaxies and thereby plays a major role in galaxy evolution. Conventionally, this multiphase gas surrounding the galaxies is expected to extend between their disks and virial radii and an interplay of baryonic matter happens between the CGM and the host galaxies. CGM gas originates from multiple physical mechanisms such as accretion from the intergalactic medium (IGM), wind from the host galaxy and gas ejected from other galaxies \citep{hafen2019}. 

The CGM has been extensively studied at low redshifts via various techniques including absorption line spectroscopy using background quasars or galaxy's own light, stacking analyses of large number of spectra, emission lines and hydrodynamic simulations \citep{cgm}. With the advent of near infrared (NIR) instruments, the CGM studies have been extended to larger redshifts, particularly making use of quasar absorption spectroscopy. The evolution of heavy elements tracing different ionisation states of the CGM is studied from the present epoch up to the end of the Epoch of Reionisation (EoR) providing valuable insights into the chemical build up and star formation rates of the Universe across cosmic time. There are many works on both low \citep{churchill, churchill2001, becker2006, becker2011, becker2015, becker2019, Zhu2013THEABSORBERS, narayanan, chen, christensen2017, Christensen2023, codoreanu, sebastian2024, abbas, churchill2025} and high \citep{tripp, tripp2008, frye2003, ryan-weber2009, cooksey2010, dodorico2010, dodorico2013, cooper, hasan, Davies2023} ionisation absorber evolution across redshift informing us about the chemical enrichment and gas phases during different periods of the timeline of the Universe. Theoretical works \citep{aguirre2001, oppenheimer2006, oppenheimer2008, finlator2008, shen2010, barai2013, keating2014, keating2016, angles2017, sorini2022, Doughty2023The5} further substantiate the idea of enrichment of diffuse gas surrounding galaxies via galactic outflows and winds and/or AGN jets and the accretion of cold gas from CGM/IGM into the galaxies; thereby influencing their subsequent star formation.

The comprehensive picture of the baryon cycle between galaxies and their surrounding diffuse media warrant the investigation of the nature of galaxies that ejected heavy elements into the CGM and the IGM. The radial distribution of absorption systems around galaxies is an efficient tool for understanding the galaxy environments at different redshifts, including metal enrichment beyond the virial radius \citep{wilde2023}. At low redshifts ($z<2$), the host galaxy-absorber pairs are broadly studied, especially for \ion{Mg}{ii} and \ion{C}{iv}. At these redshifts, the anti-correlation between \ion{Mg}{ii} absorber strength ($W$) and impact parameter ($D$) is well-established \citep{lanzetta1990, bergeron1991, bordoloi2011, Chen2010, churchill2013, nielsen2013a, nielsen2013, dutta2020, dutta2021, cherrey2025, bouche2025}. However, at $z>3$, \citet{galbiati2024}, studying 47 \ion{Mg}{ii} absorbers detected using VLT/MUSE, found that the equivalent width remains flat at large projected distances beyond the virial radii of the host galaxies and the anti-correlation does not significantly evolve compared with $z<2$ \citep{dutta2020, dutta2021, beckett2024}. Contrary to the above observation, recent work by \citet{bordoloi2023} showed that \ion{Mg}{ii} absorber strength falls off sharply with distance from the associated galaxy at $2.3<z<6.3$ using 7 \ion{Mg}{ii} systems detected in a single sightline from JWST/NIRCam. The behaviour of \ion{C}{iv} absorber strength with respect to projected distance from the host galaxy is observed to be shallower to zero anti-correlation at $z>3$ \citep{dutta2021, muzahid2021, banerjee2023, galbiati2023, beckett2024} in contrast with the strong decline observed at low redshifts \citep{bordoloi2014, burchett2016}. This could indicate a larger extent of strong \ion{C}{iv} halo in the CGM  at higher redshifts. Nevertheless, both low and high redshift works on \ion{Mg}{ii} and \ion{C}{iv} host galaxies have demonstrated that there is an enhancement in absorber strength and their detections in regions of galaxy over-densities \citep{dutta2020, dutta2021, diaz2021, galbiati2023}.

The host galaxy - absorber pair detections at high redshift not only provide insights into galaxy environment but also can be used as probes to study the ultra-violet background (UVB) towards the end of the EoR at $z\gtrsim5.3$. The luminosity function of galaxies in the early Universe can be used to understand the the contribution of ionising photons from these galaxies and constrain the process of reionisation towards its end. Cosmological hydrodynamic simulations have predicted a spatially fluctuating UVB in intensity and spectral slope after comparing with the observed absorber statistics $z\sim6$ \citep{mesinger2009, finlator2015, finlator2016, finlator2018}. The process of reionisation is inhomogeneous, occurring at denser regions of IGM where ionising sources are present and the effect of enhanced UVB in these regions can be observed in the ionisation states of the metal absorbers within the ionising bubble. Therefore, studying the metal ions and their relation with host galaxies will provide more information on the ionising photons and chemical build up in such environments.

\citet{diaz2014, diaz2015, diaz2021} searched for galaxies associated with \ion{C}{iv} systems at $z>4.7$ and detected low mass, faint Lyman $\alpha$ emitters (LAEs) postulating that these galaxies are major drivers of reionisation towards its tail end and chemical evolution of CGM and IGM. Furthermore, \citet{bielby2020, Kashino2023} found that there is an overdensity in galaxies around strong \ion{C}{iv} absorption systems detected at impact parameters that are twice the virial radii of the galaxies. Moreover, the high redshift LAEs at $z\sim3-4$ are found along gas filaments rich in \ion{H}{i} and enriched by \ion{Mg}{ii} and \ion{C}{iv} systems \citep{galbiati2023, galbiati2024}. 

The majority of the $z\sim6$ works on host galaxy - absorber connections searched for LAEs around \ion{C}{iv} systems possibly because of the higher chance of detecting galaxies in an ionised environment compared to neutral regions that are not yet ionised \citep{keating2020}. Moreover, it is easier to detect \ion{C}{iv} systems at high redshift with respect to \ion{Mg}{ii} systems. The latter falls in the near infrared region of the spectrum which is more affected by telluric or sky lines. Nevertheless, strong \ion{Mg}{ii} absorbers ($W>1.0$\AA) are considered to trace galactic outflows owing to their detection along the galaxy minor axis at $z\lesssim1$ \citep{bouche2012, kapzak2012, lan2014}. At $z>2$, these strong systems trace the trend in global star formation history, further reinstating that the low redshift strong \ion{Mg}{ii} behaviour might be reflected in the early Universe \citep{matejek, chen, sebastian2024}.

This work aims to understand galaxy environment and its evolution by searching for Lyman $\alpha$ emitting galaxies at $2.9<z<6.7$ around \ion{Mg}{ii} and \ion{C}{iv} systems that were detected in the E-XQR-30 survey \citep{xqr30catalog}. The analysis builds on previous literature by investigating \ion{Mg}{ii} and \ion{C}{iv} host galaxies at $z>3$ with a larger sample. Systematic searches for host galaxies around absorbers using multiple lines of sight at $z>5$ are still uncommon. However, the growing amount of archival data on high-redshift quasar fields (e.g. MARQUIS project, Meyer et al. in prep) and the completion of the ESO/MUSE Large Program (LP) REQUIEM ((P.I.: Emanuele Paolo Farina) now enable the search of galaxies around absorbers. The work presented here is a precursor to the forthcoming REQUIEM survey. We make use of the area and sensitivity of VLT/MUSE enabling us to detect even the faint galaxies associated with these systems. The host galaxies are blindly searched using Lyman $\alpha$ emission line in three MUSE pointings centred on three quasar lines of sight (see Section \ref{sec:methods}) and then checked for association with \ion{Mg}{ii} and \ion{C}{iv} using a velocity window of $\pm1000$ km/s from the inferred systemic redshift of the LAEs. We also obtained data from Keck/NIRC2 with the aim of calculating the stellar masses of the host galaxies. The properties of the detected LAEs are determined and analysed in connection with those of the metal absorbers along the lines of sight (see Section \ref{sec:results}). The implications of the observed galaxy-absorber connections are discussed in detail in Section \ref{sec:discussion}. 

For this work, the $\Lambda$CDM cosmology from \citet{Planck2020} with $H_0=67.7~\text{km s}^{-1}\text{Mpc}^{-1}$ and $\Omega_\text{m}=0.31$ is adopted.

\section{Methods}
\label{sec:methods}

This section outlines the different techniques employed for the detection and identification of Lyman $\alpha$ emission lines from the deep quasar fields observed using MUSE.

\subsection{MUSE Data}
\label{sec:data reduction} 

Data required for detecting galaxies associated with \ion{Mg}{ii} and \ion{C}{iv} are obtained from ESO archive compilation for deep VLT/MUSE observations called MARQUIS (Meyer et al. in prep). Three quasar fields at $z\sim6$ - SDSSJ1030+0524, SDSSJ1306+0356 and PSOJ158-14 - are taken from MARQUIS for this work which have been observed for at least 6 hours, such that even $\text{sub-L}_{\text{Ly}\alpha}^*$ ($<10^{42.7}~\text{ergs}^{-1}\text{cm}^{-2}$) galaxies can be detected. None of these observings have utilised the adaptive optics facility (AOF). These quasars have been observed in the Enlarged Ultimate XSHOOTER Legacy Survey of Quasars (E-XQR-30) which is a homogeneous sample of high quality $z\sim6$ quasar spectra \citep{xqr30}. The metal absorber catalogue produced from this survey \citep{xqr30catalog} is used to study the evolution of line densities of metal absorbers at $2<z<6$ \citep{Davies2023, sebastian2024} and the galaxies detected in this work are cross-correlated with the absorbers in the catalogue, in particular, \ion{Mg}{ii} and \ion{C}{iv}. Table \ref{tab:qso} shows details of the quasar fields including their coordinates, redshift, exposure time and the number of LAEs detected in each field. It should be noted that although J158 has the second highest exposure time, the observing conditions were significantly worse than J1030, resulting in higher noise in the cube compared to J1030 (for details refer to Meyer et al. in prep). Based on the minimum LAE flux that were detected from each cube, J1030 has the best signal-to-noise ratio (S/N) compared to the other fields. 
\begin{table*}
    \centering
    \caption{The quasar fields from the MARQUIS survey that are studied in this work. The quasar coordinates (RA and Dec), emission redshift, exposure time (in seconds), median seeing and median airmass of the MUSE datacubes, the corresponding observing program IDs and the number of LAEs detected in each quasar field are listed here in decreasing order of their exposure times. The minimum LAE flux that can be detected from each cube is also shown here.}
    \begin{tabular}{cccccccccc}
    \hline
        QSO & RA & Dec  & z & Exp time  & Median & Median  & Program & \#LAEs & Flux limits\\
            &  &  &    &  & seeing & airmass & ID &  & log $F$ \\
            & (hh:mm:ss) & (dd:mm:ss) & & (s) & & & & & $[\text{ergs}^{-1}\text{cm}^{-2}]$\\
        \hline
        SDSSJ1306+0356 & 13:06:08.2584 & +03:56:26.3004 & 6.033 & 38280 & 0.74 & 1.3 & 0103.A-0140(A) & 59 & -17.77\\
        PSOJ158-14 & 10:34:46.50936 & -14:25:15.8844 & 6.0685 & 30540 & 0.89 & 1.2 & 106.215A.001 & 26 & -17.79\\
        SDSSJ1030+0524 & 10:30:27.0984 & +05:24:55.0008 & 6.304 & 23152 & 0.86 & 1.2 & 095.A-0714(A) & 71 & -17.64\\
        
    \hline    
    \end{tabular}
    
    \label{tab:qso}
\end{table*}

Data from MUSE deep fields have been reduced using the default MUSE pipeline v2.8.3 using the master calibration files produced in Phase 3 available through the ESO archive and further processed using the \textit{Zurich Atmosphere Purge} \citep[ZAP,][]{ZAP} which improves sky subtraction. 

\subsection{LSDCat: Producing emission lines catalogue}
\label{sec:lsdcat}

Potential sources for Lyman $\alpha$ emission are detected using LSDCat \citep{LSDCat1.0, LSDCat2.0}. LSDCat is a Python based package particularly developed for wide-field integral spectroscopic datacubes, like MUSE cubes, to detect emission lines in them in an efficient manner. 
LSDCat has been successful in detecting Lyman $\alpha$ emission lines at high redshift from MUSE wide-field and deep-field surveys \citep[e.g.,][]{lsdcat5, lsdcat6, lsdcat4, lsdcat3, lsdcat2, lsdcat1}. In this work, we used the updated version of the algorithm called LSDCat2.0 \citep{LSDCat2.0}. A detailed description of the various algorithms used in LSDCat can be found in \citet{LSDCat1.0, LSDCat2.0}.

For this work, bright continuum subtracted MUSE cubes via median filtering along the spectral direction are used. The bright sources that were not removed through median filtering are masked along with edges of the cube. For 3D matched filtering, default values of the requires parameters are applied. A circular symmetric Gaussian point spread function (PSF) is approximated for the spatial profile. Given that we search for compact emission lines, the wavelength dependence of the seeing is neglected here by setting the polynomial coefficient to the default value of 0.8. Setting the polynomial of the wavelength dependence of the seeing as a constant will result only in a very modest reduction in the sensitivity at the lowest and highest wavelengths \citep{LSDCat1.0}. For this work, exposure maps are not used for the spectral filtering as they were consistent across the spaxels, excluding the masked edges. We also input a 1D variance spectrum averaged over the variance HDU. The default detection threshold value of 8.0 is used for detecting the emission lines. LSDCat creates an intermediate catalogue with all detections with a S/N above the detection threshold. See Appendix \ref{sec:LAE detction flowchart} for more details. The intermediate catalogue is then passed as input for producing the final catalogue.

For creating the final catalogue consisting of the parameters for each emission line, an analysis threshold of 3.5 is used in this work. This threshold does not affect the number of detected sources in the previous step but rather, determines the S/N of the voxels around a source that must be included when calculating their fluxes \citep{LSDCat1.0}. LSDCat calculates flux by summing the pixels from narrow band images, whose spectral boundaries are determined by the analysis threshold of 3.5. The flux is measured in circular apertures whose radii is equal to multiples $k$ of Kron radius \citep{kron1980} for a detected line. 3 Kron radii encompasses $>99\%$ of flux for Gaussian profiles \citep{graham2005}. Moreover, in this step, LSDCat calculates 3D S/N weighted centroid positions of the detected sources based on the analysis threshold. The details of various other measurements available in the final catalogue can be found in \citet{LSDCat1.0}. 

\subsection{QtClassify: Identifying LAEs}
\label{sec:qtclassify}
QtClassify \citep{qtclassify} is a graphical user interface (GUI) that helps classify candidate emission lines detected in MUSE datacubes with the help of software like LSDCat and determine their redshifts. 

\begin{figure}
    \centering
    \includegraphics[width=\linewidth]{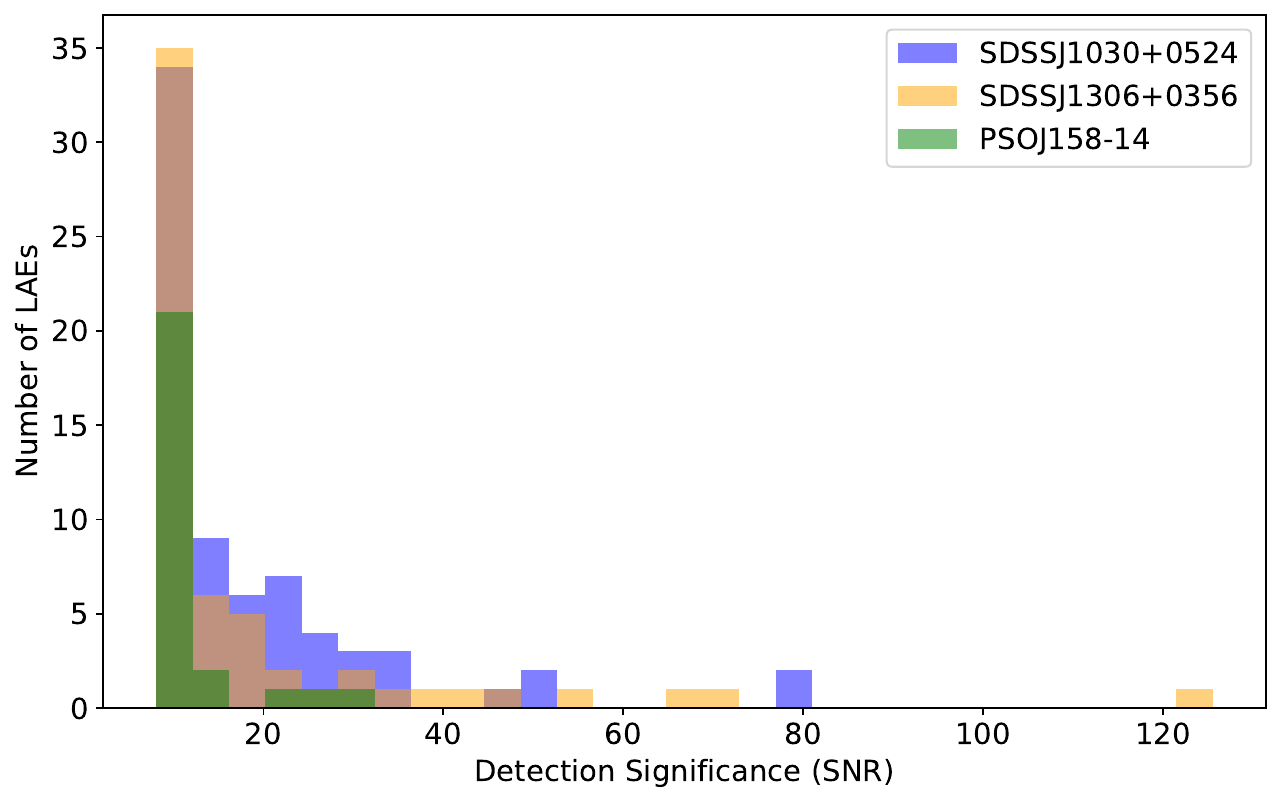}
    \caption{The detection significance of the LAEs detected in all the three fields. Histograms for detections from each field is colour-coded accordingly.}
    \label{fig:lae det sig}
\end{figure}

In this work, we detected LAEs with detection significance (S/N) ranging from 8 to 80 in J1030, 8 to 126 in J1306 and 8 to 30 in J158. Surprisingly, the LAEs detected in J158 have the lowest S/N although it has a high exposure time (see Table \ref{tab:qso}). Figure \ref{fig:lae det sig} shows the detection significance histograms for LAEs detected in each quasar field. It is evident that for all the three fields, the number of detections with the least confidence is the highest relative to higher confidence LAEs.

A flowchart showing the different steps followed in detecting LAEs from MUSE datacubes using LSDCat and QtClassify - from the input cube to the generation of the output catalogue consisting of the classification of detected emission lines and their properties - is given in Figure \ref{fig:LAE workflow}. The emission lines are classified individually by the first and the second authors of this paper and then the classifications that differ in their individual efforts are re-scrutinised together for robust identifications. In this endeavour, lower redshift contaminants are removed when multiple emission lines are detected and can be associated with \ion{O}{iii}, \ion{H}{$\alpha$}, etc. and the \ion{O}{ii} doublet. We paid particular attention to doublet candidates at wavelengths $>7122$\AA~by conservatively labelling any double peak sources matching the \ion{O}{ii} separation as \ion{O}{ii} emitters. Due to the wavelength of \ion{O}{ii} emission ($\lambda\lambda$3727, 3729\AA), such sources have too low redshift to overlap with \ion{Mg}{ii} absorbers in E-XQR-30 and are therefore, not considered in the analysis.

At the end of this process, we identified 72 LAEs in J1030, 63 in J1306 and 27 in J158.

\subsection{pyPlatefit: Measuring Lyman alpha emission line flux}
\label{sec:pyplatefit}
After using QtClassify to classify the emission lines detected by LSDCat, the flux of each Lyman alpha emission line is calculated with the help of pyPlatefit \citep{pyplatefit}. This is a python module, primarily using the non-linear least square minimisation technique for fitting emission and absorption lines, particularly developed for MUSE datacubes. 

For the Lyman $\alpha$ line, an asymmetric Gaussian function is used which is modelled using four parameters: central wavelength ($\lambda_0$), wavelength dispersion ($\sigma$), peak flux ($F_0$) and skewness ($\gamma$). The skewed Gaussian used to fit Lyman $\alpha$ is given as 
\begin{equation}
\label{eq:gaussian}
    F(\lambda) = F_0\bigg[1+\text{erf}\bigg(\gamma\frac{\lambda-\lambda_0}{\sqrt{2}\sigma}\bigg)\bigg]\text{exp}\bigg(-\frac{(\lambda-\lambda_0)^2}{2\sigma^2}\bigg)
\end{equation} For a double peaked Lyman $\alpha$, the flux is calculated as a sum of two asymmetric Gaussian functions. 

The input spectrum for each source is obtained from the median filtered datacube (see \ref{sec:lsdcat}) for each quasar field using an aperture size corresponding to the 3 Kron radius calculated by LSDCat. For the majority of sources, the 3 Kron radius corresponds to $1.8"$. A flowchart showing the steps followed in fitting the Lyman $\alpha$ emission lines in pyPlatefit is shown in Figure \ref{fig:pyplatefit}.

\begin{figure}

\begin{subfigure}{0.5\textwidth}
\includegraphics[width=0.9\columnwidth]{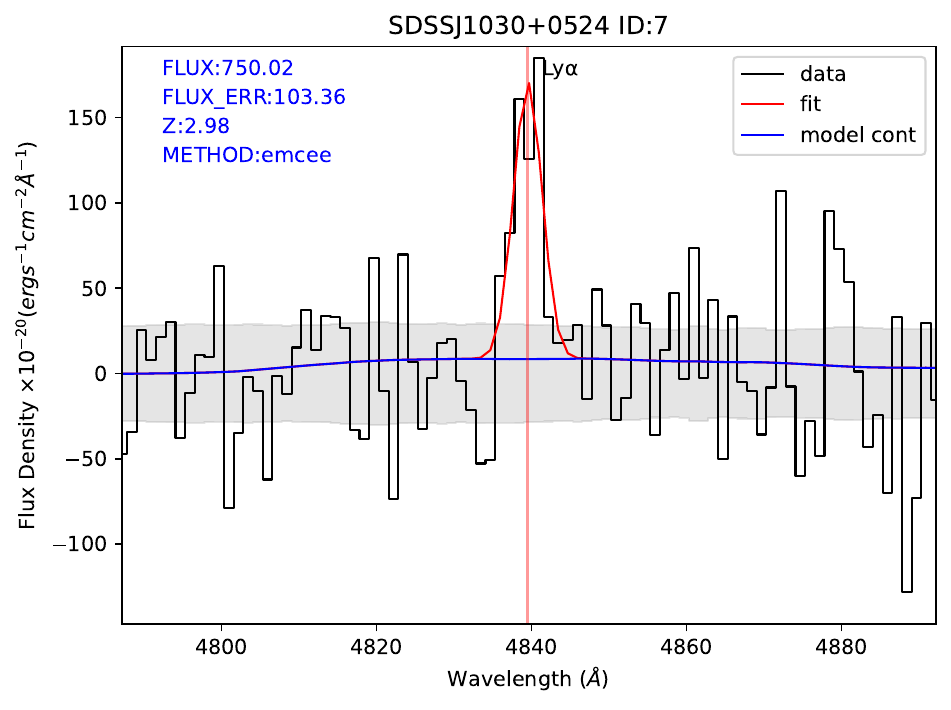} 
\label{fig:single fit}
\end{subfigure}
\begin{subfigure}{0.5\textwidth}
\includegraphics[width=0.9\columnwidth]{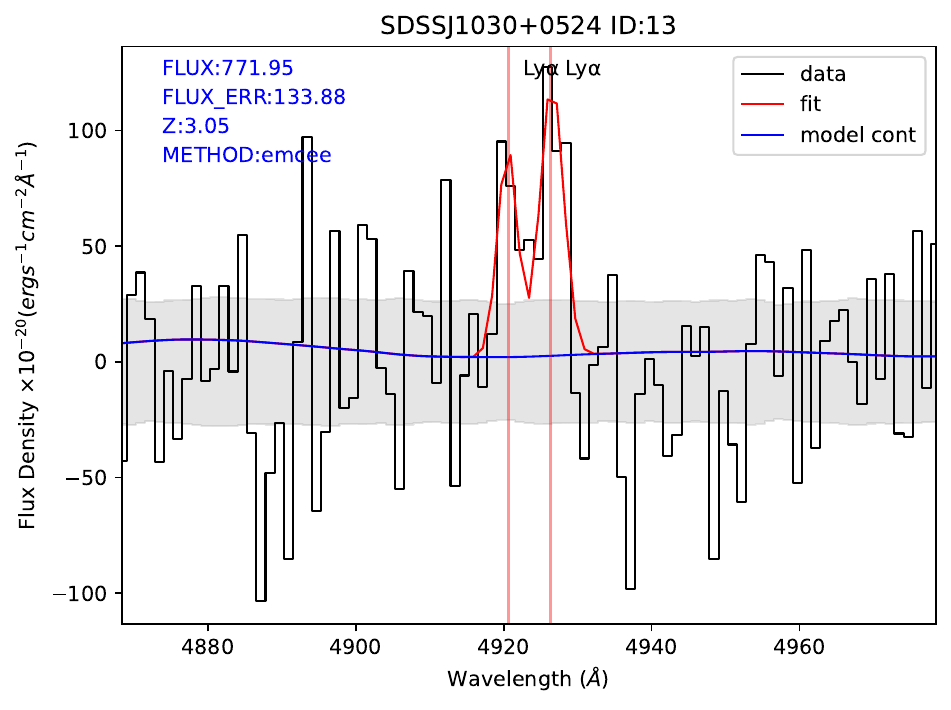}
\label{fig:double fit}
\end{subfigure}

\caption{The Gaussian fits produced by pyPlatefit for two LAEs in J1030. The redshift, flux, the error in flux and the method adopted for error estimation are annotated in the respective figures. The spectrum is shown in black, the continuum fit in blue and the fitted profile is shown in red. The grey shaded region marks the noise associated with the spectrum. The top panel shows a single asymmetric Gaussian profile for the source with object ID 7 while the bottom panel shows a doublet fit for object ID 13 in the same quasar field. For doublet line profiles similar to ID 13, the observed peak separation is not consistent with low redshift \ion{O}{ii} doublets.}
\label{fig:fits}
\end{figure}

Examples of single and doublet profile fits of Lyman $\alpha$ lines in J1030 performed by pyPlatefit can be seen in Figure \ref{fig:fits}. The spectrum is shown in black, the continuum fit in blue and the line fit in red. The flux, error associated with flux (both in units of $10^{-20}\text{ergs}^{-1}\text{cm}^{-2}$), the redshift and the method of error estimation for the emission lines are marked in the respective figures.
A reasonable fit will have more than one channel, significantly above the noise shown in gray and significant compared to surrounding noise features.

After the emission line fitting procedure using pyPlatefit, the number of LAEs reduced to 71 in J1030, 59 in J1306 and 26 in J158. Table \ref{tab:LAE detections} provide the details of the LAEs detected in this work. This includes the quasar field from which they are detected, the identifiers assigned to them by LSDCat, the sky coordinates, the Lyman $\alpha$ emission ($z_\text{Lya}$) and systemic ($z_\text{systemic}$) redshifts, the observed full widths at half maximum (FWHM) in units of \AA, the flux and associated errors measured using PyPlatefit and flags for doublet profiles. The final column consists of cross-matched detections from \citet{diaz2021} for \ion{C}{iv} associated LAEs and \ion{O}{iii} emitters reported by \citet{Kashino2025} in J1030. The various analyses performed on the detected LAEs  and the implications of the results obtained, are discussed in the upcoming sections.

\subsection{Systemic redshift of LAEs}
\label{subsec:systemic z}
Due to the resonant scattering of Lyman $\alpha$ photons within the neutral gas in the interstellar medium and in the vicinity of galaxies, centre of the single peaked Lyman $\alpha$ line does not trace the systemic redshift of the LAEs. Accurate redshifts are obtained by detecting other nebular lines from the galaxies \citep{mclinden2011, song2014}, absorption lines around the galaxies \citep{rakic2011} or by modelling the Lyman $\alpha$ profile \citep{hashimoto2015}. Being a survey of high redshift galaxies emitting Lyman $\alpha$ photons, it is necessary to recover the systemic redshift of the LAEs to avoid incorrect results on the various analyses performed on this sample where accurate redshift measurements are required. Due to the limits in the spectral coverage of MUSE data, we do not have access to other emission lines from LAEs and therefore, use the tight correlations obtained in \citet{Verhamme2018} between the velocity offset of the red peak of Lyman $\alpha$ line and half of the peak separation for double peaked LAEs and the FWHM for single peaked LAEs for a redshift across $3<z<6$. Using the above correlations, the systemic redshift of the LAEs can be retrieved with an uncertainty lower than $\pm100~\text{km s}^{-1}$. Based on this technique, the Lyman $\alpha$ velocity offset from the systemic redshift ranges from 0 to $\approx 761 \text{ km s}^{-1}$ for the LAEs in our sample. The LAE redshifts mentioned in this work are the systemic redshifts unless otherwise specified.

\subsection{Cross-validation with literature}
\label{subsec:cross validation}
Since the MUSE datacubes originate from the public ESO archive, we can validate our LAE detection methods by comparing our results with existing literature. Specifically, \citet{diaz2021} conducted a targeted search for LAEs associated with \ion{C}{iv} systems in the J1030 quasar field, identifying eight sources. In this work, we performed a blind search for LAEs of the same field. By cross-matching our detections with those reported by \citet{diaz2021}, we found that our method successfully recovered all eight of their previously identified LAEs. However, we discovered 18 LAEs instead of 8 around the 4 \ion{C}{iv} systems in J1030 studied in \citet{diaz2021} due to the ZAP reduced cube used in this work. We also found additional LAEs associated with two \ion{C}{iv} systems at $z=5.120$ and $z=5.126$ towards J1030, newly reported in the E-XQR-30 absorber catalogue \citep{xqr30catalog}. Table \ref{tab:LAE detections} includes a column showing \citet{diaz2021} LAE IDs detected in both works.

The discovery of an LAE at $z=6.03$ at a projected distance of $~29$ pkpc has been reported by \citet{durovcikova2025} in the quasar field J158 using data from MUSE. This LAE is located at the outer edge of the quasar proximity zone and is assumed to be associated with an extremely metal poor absorption system. The detected LAE has a flux of $\text{log }F\approx-17.69$ which is too faint to be detected in this work (see Table \ref{tab:arctan fit}). 

Recent work by \citet{Kashino2025}, using \textit{JWST} EIGER data, detected 948 \ion{O}{iii} emitting galaxies across $ 5.33<z<6.97$ along six quasar sightlines including J1030. There are 24 LAEs within the redshift range of \ion{O}{iii} emitters detected in J1030. Comparing the \ion{O}{iii} emitters from \citet{Kashino2025} with our LAE detections in J1030, 8 LAEs are confirmed to have \ion{O}{iii} emission across $5.5<z<5.8$ as can be seen in Table \ref{tab:LAE detections}. Our LAE detections at $z>5.5$ are consistent with the enhanced Lyman $\alpha$ transmission at $z>5.5$ due to the combined effect of local radiation fields around galaxies and IGM density \citep{Kashino2025}. However, we do not detect any matching \ion{O}{iii} emitters for LAEs beyond $z\sim5.8$.  

\subsection{NIRC2 Data}
\label{subsec:NIRC2}
We obtained H and Kp band images of the J1030 field on 10 Jan 2025 using Keck/NIRC2 with NGS-AO to estimate the stellar masses of LAEs detected in the field. We used the wide-field mode, with a 40x40 arcsec field-of-view and 0.04" pixels, centred near the quasar at PA=0 deg such that the images cover the central ~44\% of the MUSE field. Individual images had an exposure time of 50s and we adopted the standard 3-by-3 point dither patten, reaching a total on-source integration time of 3000s in K$_\text{p}$ and 2400s in H. The data were reduced using custom algorithms which applied flat-fielding, masked bad pixels, removed cosmic rays, performed background subtraction by median-combining frames at different dither positions, de-warped frames using the method described on the NIRC2 astrometry page \footnote{\url{https://www2.keck.hawaii.edu/inst/nirc2/dewarp.html}}, and stacked all the images. Flux calibration was performed using standard star observations. We measured zero-points of 24.2 in K$_\text{p}$ and 24.8 in H.

\begin{figure}
    \centering
    \includegraphics[width=\linewidth]{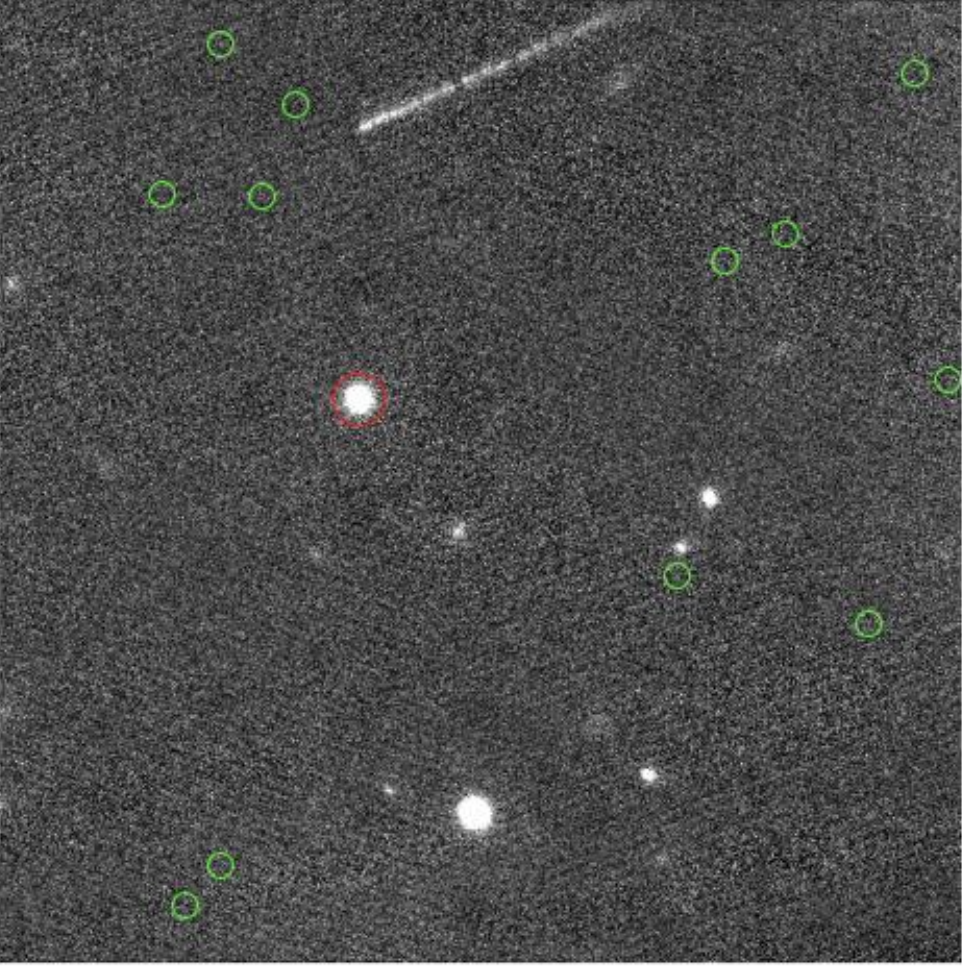}
    \caption{The Keck/NIRC2 image using H filter with the LAEs (green) and quasar (red) overlaid on them. No detections can be found for the LAEs.}
    \label{fig:keck images}
\end{figure}

Given that the lower redshift LAEs will be brighter and thus, more likely to be detected, further analysis on LAEs detected in this work is restricted to the redshift window of $3<z<4$. On careful inspection of the images from H and K$_\text{p}$, no detections were found corresponding to the LAE coordinates from MUSE datacubes (see Figure \ref{fig:keck images}). We are, therefore, only able to compute upper limits on their stellar masses (see Section \ref{subsec: stellar masses}). 

\section{Results}
\label{sec:results}

\subsection{LAE distribution}
\label{subsec:lae distribution}
There are 71 confirmed LAEs in J1030, 59 in J1306 and 26 in J158 at $2.9<z<6.7$. The distribution of these associated galaxies with respect to various parameters will help in understanding the properties of the detected LAEs in the quasar fields. 

\begin{figure}

\begin{subfigure}{\columnwidth}
\includegraphics[width=\linewidth]{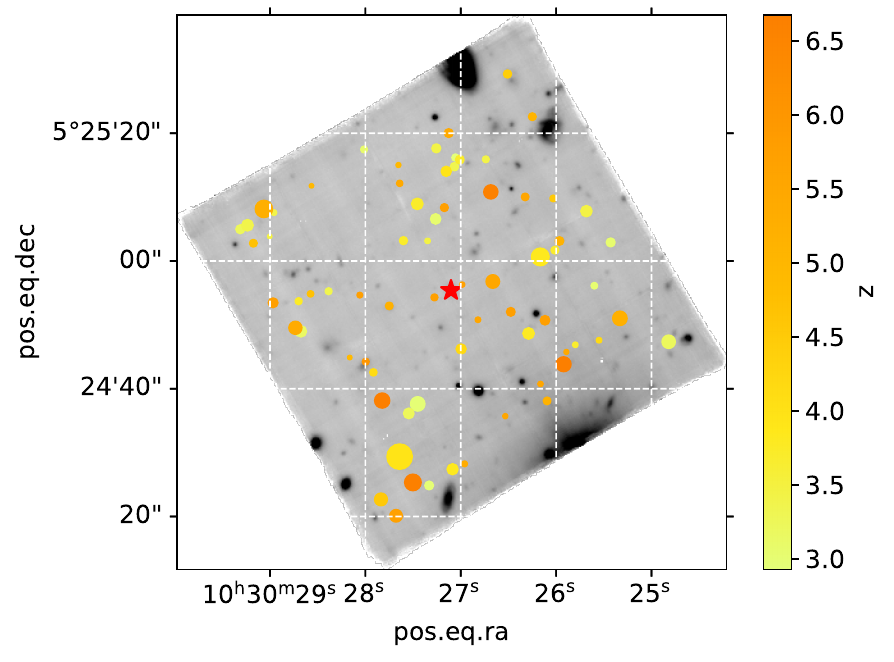} 
\caption{J1030}
\end{subfigure}
\begin{subfigure}{\columnwidth}
\includegraphics[width=\linewidth]{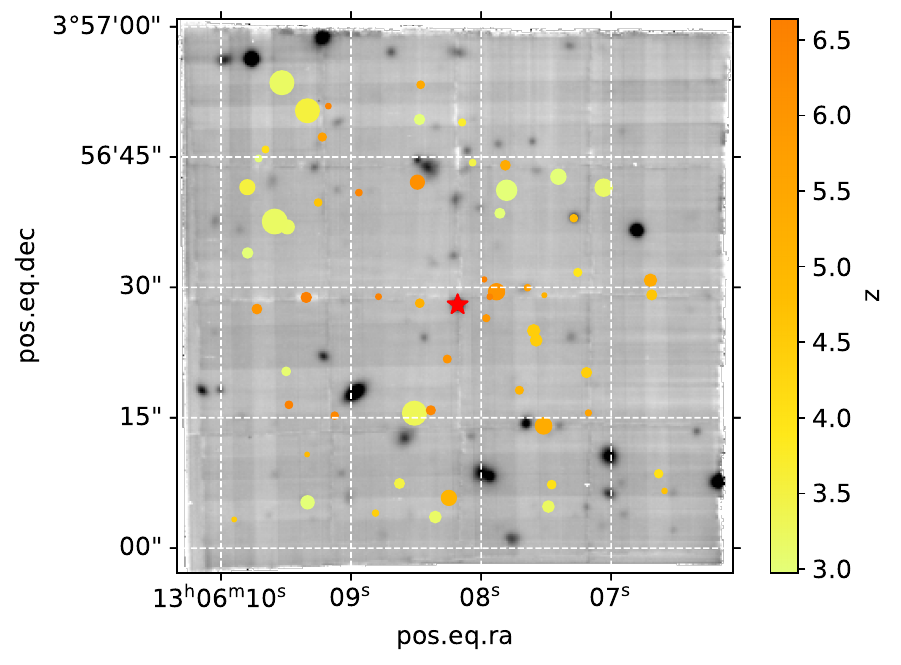}
\caption{J1306}
\end{subfigure}
\begin{subfigure}{\columnwidth}
\includegraphics[width=\linewidth]{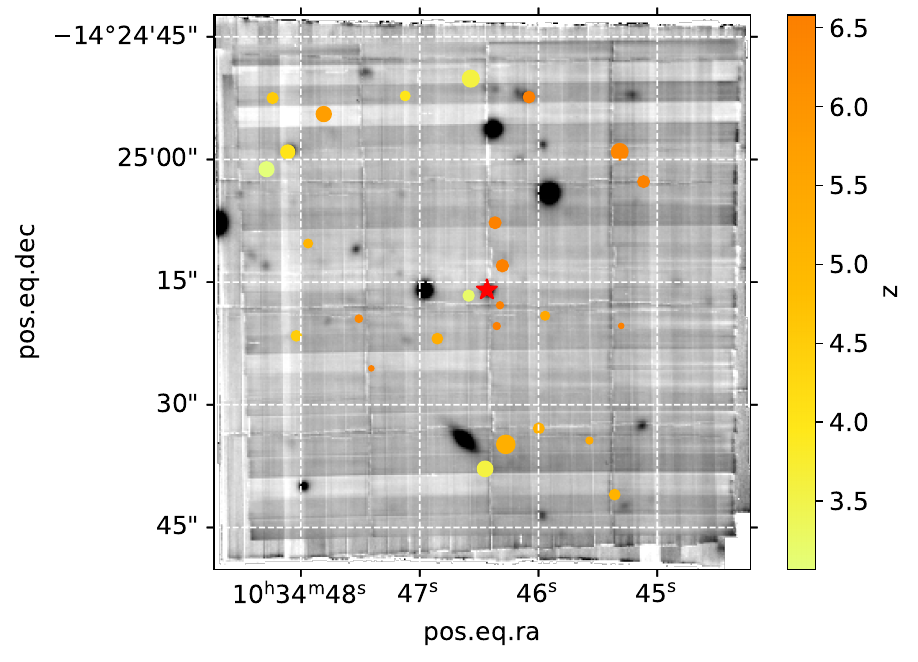} 
\caption{J158}
\end{subfigure}

\caption{The spatial distribution of LAEs detected in each quasar field. The detections in each cube are shown using the solid dots from top to bottom respectively. The LAEs are colour coded based on their systemic redshifts and the marker size depends on the flux measured from these sources. The central quasar is marked using a red star.}
\label{fig:detections}
\end{figure}

Figure \ref{fig:detections} shows the LAEs detected in each quasar field as solid dots colour coded based on their redshifts. The marker size is scaled depending on their line flux values. The quasar is indicated using a red star in each field. From top to bottom, the detections in J1030, J1306 and J158 are depicted respectively. For J1030, the highest number of LAEs are detected at $2.9<z<3.9$ and $5.3<z<5.7$ while for J1306, the most number of LAEs are detected at $2.9<z<3.4$. In J158, most LAEs are present at $5.9<z<6.6$, beyond the quasar redshift. The LAE with the highest emission of $\text{log}~F[\text{ergs}^{-1}\text{cm}^2]=-16.2$ is found in J1030 at $z=3.9$ while the LAE with lowest emission of log $F[\text{ergs}^{-1}\text{cm}^2]=-17.8$ at $z=4.9$ belongs to J1306. 

\subsection{Absorber-centric analysis}
\label{subsec:absorber-centric analysis}
\begin{figure*}

\begin{subfigure}{\columnwidth}
\includegraphics[width=\linewidth]{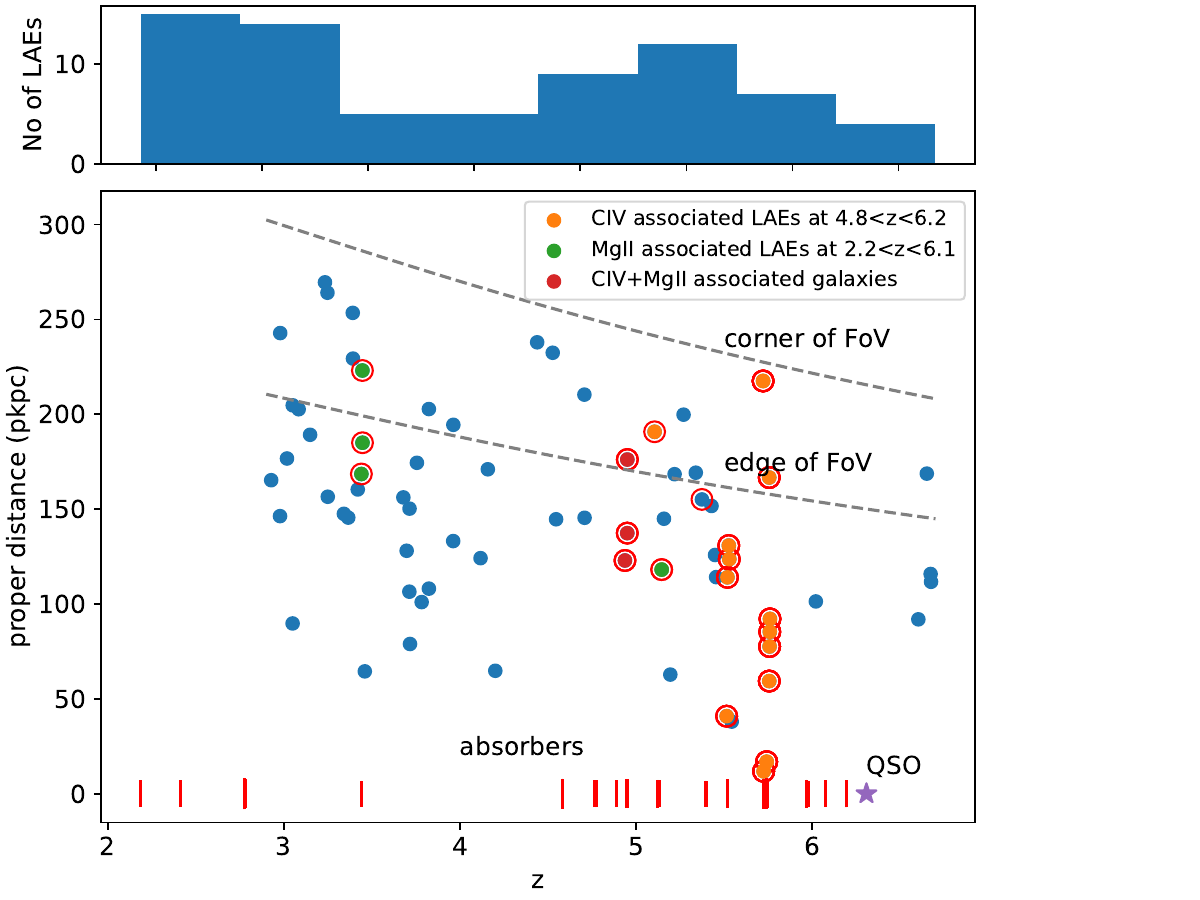} 
\caption{J1030}
\end{subfigure}
\begin{subfigure}{\columnwidth}
\includegraphics[width=\linewidth]{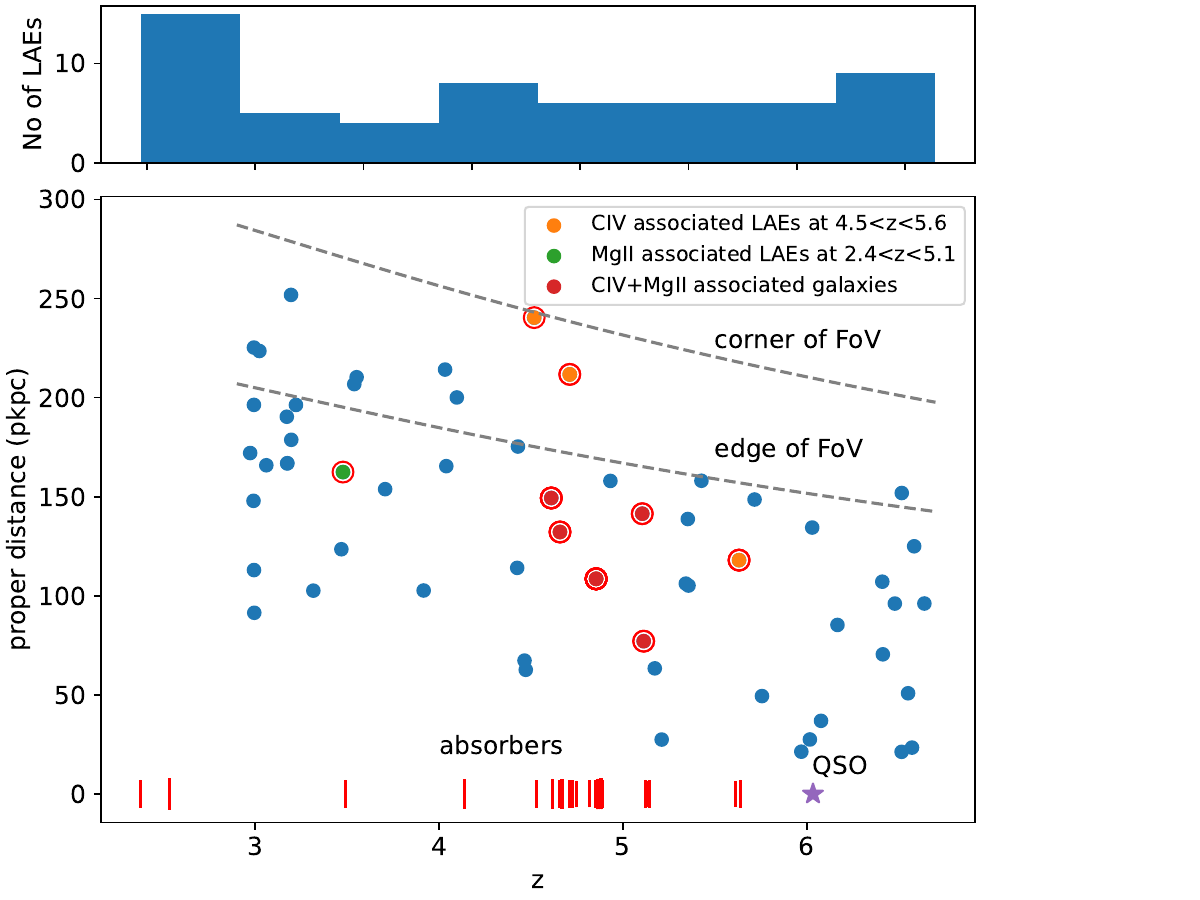}
\caption{J1306}
\end{subfigure}
\begin{subfigure}{\columnwidth}
\includegraphics[width=\linewidth]{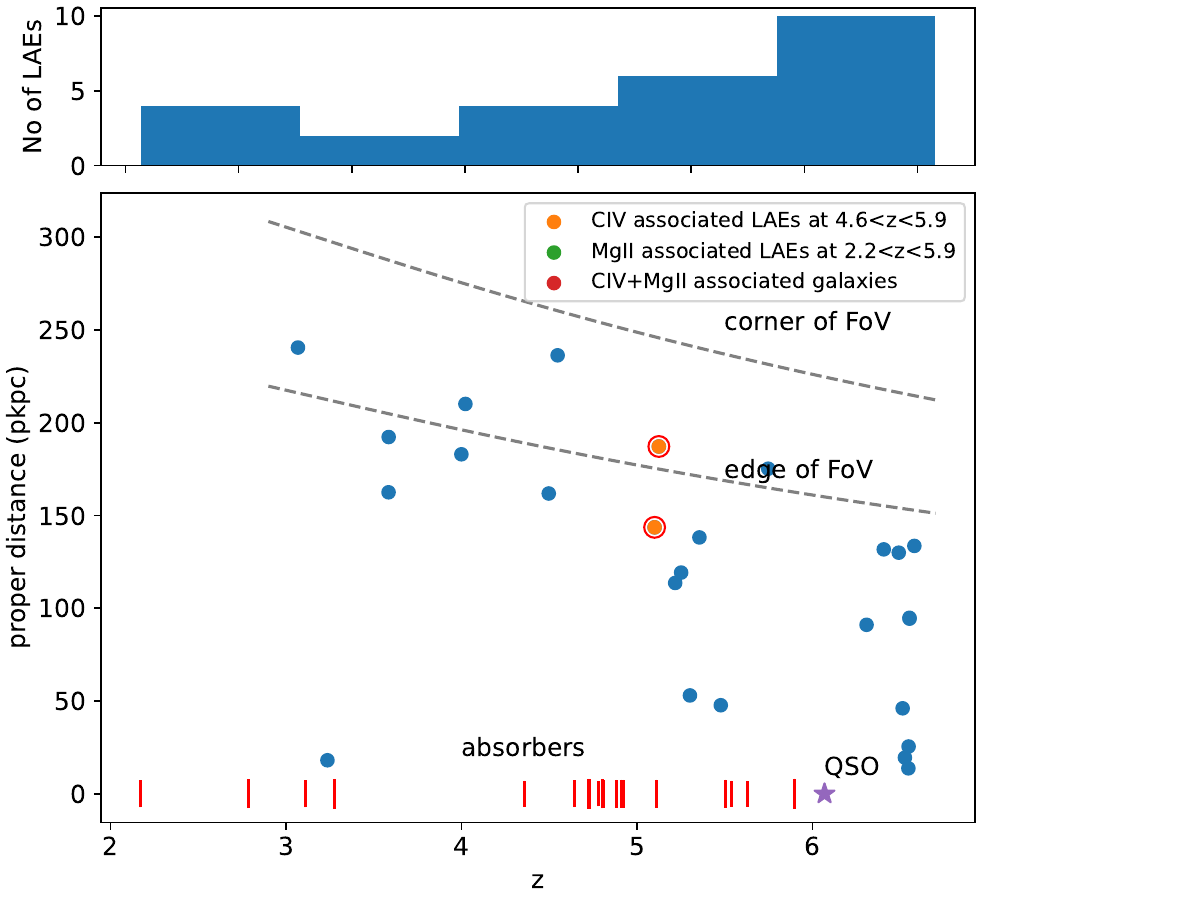} 
\caption{J158}
\end{subfigure}

\caption{The projected distance of the detected LAEs from the quasar sightline as a function of redshift. The LAEs are indicated using blue dots. Galaxies that are located at a velocity separation of $\pm1000\text{ km s}^{-1}$ from the sightline are highlighted in red circles. Among them \ion{C}{iv}-only and \ion{Mg}{ii}-only associated LAEs are shown in orange and green colours respectively. LAEs that are associated with both \ion{C}{iv} and \ion{Mg}{ii} are indicated in red. The absorbers from E-XQR-30 along the line of sight are shown using red vertical lines corresponding to their redshifts and the background quasar using a purple star. The redshift ranges at which \ion{Mg}{ii} and \ion{C}{iv} can be detected in the corresponding E-XQR-30 quasar spectra is given in the legend. The top panel gives the redshift distribution of LAEs detected in each quasar field. }
\label{fig:lae_proj_dist}
\end{figure*}

Identifying the galaxies associated with the metal absorption systems detected in the quasar sightline is crucial to understand the nature of the galaxies that enriched the Universe with heavy elements. Absorber-centric analyses result in unbiased selection of associated galaxies independent of their luminosity or other criteria used for pre-selecting galaxies. The projected distance of LAEs from the quasar line of sight as a function of redshift can be viewed in Figure \ref{fig:lae_proj_dist}. The LAEs are marked using blue dots and the absorbers detected in E-XQR-30 along the sightline are marked using red vertical lines at their respective redshifts with the host quasar denoted as the purple star. The field of view (FoV) for MUSE is marked using the grey dashed lines indicating that only LAEs lying up to a projected distance of $\sim300$ pkpc and between $2.9<z<6.6$ can be detected. At $z<5$, no galaxies are found within 50 pkpc of the line of sight. To understand if this gap can be accounted for by the expansion of the Universe, we looked at the comoving transverse distance between the galaxies and the sightline and found that the lack of galaxies close to the sightline at low redshifts is not statistically significant. A histogram showing the redshift distribution can be viewed at the top panels of each figure. The peaks at certain redshift ranges in each quasar fields might be due to clustering of galaxies. 

\begin{figure}
    \centering
    \includegraphics[width=\linewidth]{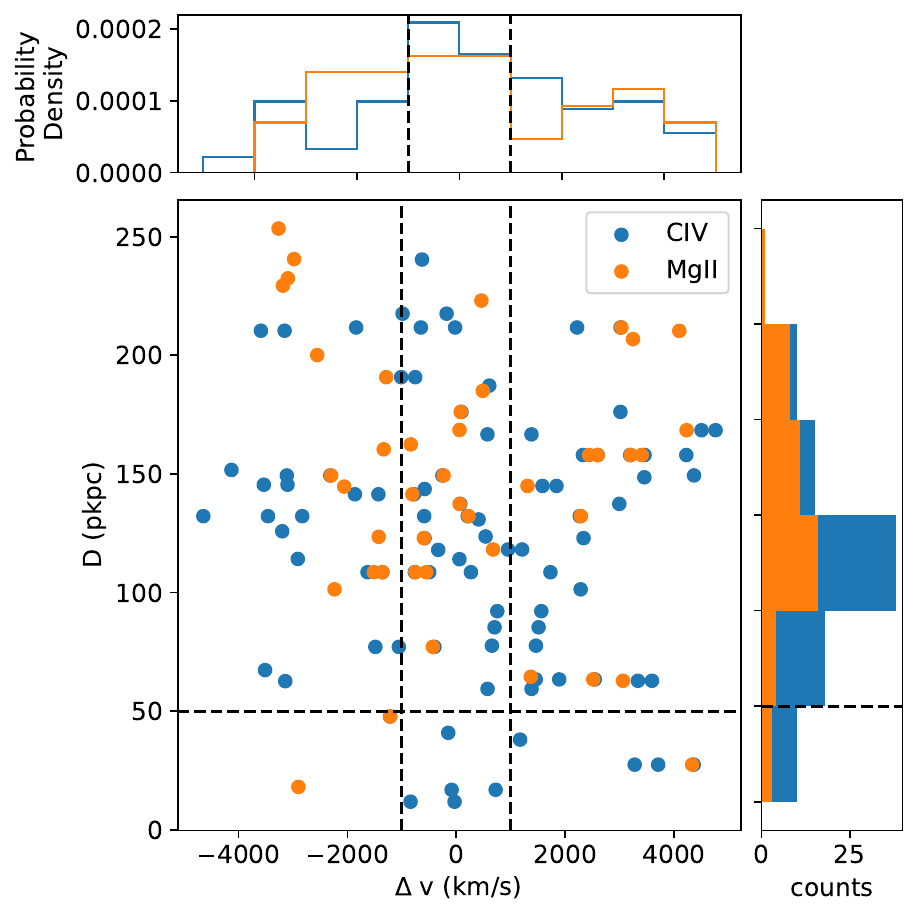}
    \caption{The scatter plot of velocity separation ($\Delta v$) between galaxies and absorbers within $\pm5000~\text{km s}^{-1}$ and their impact parameters ($D$). The blue colour represent \ion{C}{iv} absorber-galaxy separations while the orange colour represents the \ion{Mg}{ii}-galaxy separations. We define as associated the absorber-galaxy pairs with $\Delta v<\pm1000~\text{km s}^{-1}$. The top panel shows the distribution (normalised to 1) of velocity separation with a relatively high peak within $+/-1000~\text{km s}^{-1}$. The histogram on the right represents the distribution of impact parameters. Most of the \ion{C}{iv} and \ion{Mg}{ii} absorbers are located within 100-150 pkpc from the LAEs.}
    \label{fig:dv versus D}
\end{figure}

We adopt an empirical approach by inspecting the peak in the $\Delta v-D$ distribution over a window of $\pm5000~\text{km s}^{-1}$ (Figure \ref{fig:dv versus D}) and choose $\pm1000~\text{km s}^{-1}$ and $<250$ pkpc separation as the definition of an absorber-galaxy association in this work. Such an association includes absorbers arising in the halo of an immediate galaxy or a local galaxy overdensity and is similar to that adopted by other high-redshift galaxy-absorber studies \citep{mackenzie2019, bielby2020, Kashino2023, Zou2024AVLT}. However, no peak is observed at $\Delta v\sim0$ within 50 pkpc in Figure \ref{fig:dv versus D} and the physical interpretations for the absence of galaxies at small velocity separations and impact parameters are discussed in detail in Section \ref{subsec:flat W-D}.
If the systemic redshift(s) of a galaxy/galaxies are located at $\leq\pm1000$ km/s with respect to the absorber redshift, then we assume that the LAE/LAEs are  associated with the absorber. The velocity separation between the LAEs in our sample and the E-XQR-30 absorbers for the three sightlines are calculated and the associated galaxy/ies within $\pm 1000\text{ km s}^{-1}$  are highlighted in red as shown in Figure \ref{fig:lae_proj_dist}. 

Among the 32 associated galaxies, 18 LAEs are exclusively associated with \ion{C}{iv} as shown by orange points while 5 LAEs are exclusively connected with \ion{Mg}{ii} which are represented by green colour in all the three quasar fields.  There are 8 LAEs with both \ion{C}{iv} and \ion{Mg}{ii} absorbers around them within $\pm1000~\text{km s}^{-1}$ as indicated by the red dots in Figure \ref{fig:lae_proj_dist}. 

Focusing on the most common ions in the three quasar fields, \ion{Mg}{ii} and \ion{C}{iv}, 18/39 ($\sim49$\%) \ion{C}{iv} systems are observed to have associated galaxies and 9/22 ($\sim 41$\%) \ion{Mg}{ii} systems have associated LAEs within a proper distance of $\sim 250$ kpc across 3.4<z<5.8. Among them, 3 \ion{Mg}{ii} and 6 \ion{C}{iv} have multiple galaxies associated with them within $\pm 1000 \text{ km s}^{-1}$. 

Although \ion{C}{iv} is expected to be found in regions of galaxy overdensity \citep{adelberger2005, Kashino2023}, the non-detections of associated galaxies around 50\% of \ion{C}{iv} may be accounted to the galaxies being fainter than the detection limit of this work. \citet{diaz2021} observed that all \ion{C}{iv} absorbers with log $N[cm^{-2}]>13.5$ in J1030 have at least one galaxy association. Extending the analysis to the three quasar fields, we find that majority of strong \ion{C}{iv} systems have associated LAEs in our sample but there are exceptions where no galaxies are found around a few of these strong systems. Out of 18 strong \ion{C}{iv} systems in the three sightlines, only 11 of them have at least one galaxy within 250 pkpc.

\citet{sebastian2024} observed that strong \ion{Mg}{ii} systems ($W>1.0\text{\AA}$) trace the declining trend in global star formation history across cosmic time and therefore, expected these absorbers to have associated galaxies associated with them within $\sim250$ pkpc. Surprisingly, only 1/5 of the strong \ion{Mg}{ii} systems are observed to have LAEs in their near vicinity. Among them, only one strong \ion{Mg}{ii} system which also has \ion{C}{iv}, is found to have galaxy associations. The rest of the systems might be residing in dense or dusty regions where the Lyman $\alpha$ emission from the galaxies is attenuated (see Section \ref{subsec:flat W-D}). Remarkably, among weak \ion{Mg}{ii} absorbers ($W<0.3\text{\AA}$), 5/12 have galaxies associated with them at $\lesssim250$ kpc but these systems were expected to be farther out in the CGM regions compared to strong absorbers, based on the anti-correlation between equivalent width and impact parameter observed in \citet{bordoloi2023} at $2.3<z<6.3$. 

For \ion{Mg}{ii}, the associated galaxies are located from $\sim$77 to $\sim$223 pkpc and for \ion{C}{iv}, the impact parameter ranges between $\sim12$ to $\sim240$ pkpc. Figure \ref{fig:dv versus D} shows the distribution of impact parameters from the three quasar fields for \ion{C}{iv} (upper panel) and \ion{Mg}{ii} (lower panel). The impact parameters calculated for the LAEs in each quasar field are colour coded accordingly. The histograms indicate that most of the \ion{Mg}{ii} and \ion{C}{iv} absorbers are located at an impact parameter of 100-150 kpc. It should be noted here that the concentric annuli of radial distance from a galaxy cover the same area. 

The typical virial radii for galaxies in our sample where the mean density of the galaxy reaches 200 times the critical density of the Universe can be calculated using the following equation
\begin{equation}
\label{eq:virial radius}
    R_{vir}^3 =\frac{M_{halo}G}{100H^2(z)}
\end{equation} where $M_{halo}$ is the halo mass, $G$ is the gravitational constant and $H(z)$ is the Hubble parameter at redshift $z$. Here, we adopt the dark matter halo mass $\text{log}(M_{halo}/[h^{-1}M_\odot])=11.34_{-0.27}^{+0.23}$ from \citet{lsdcat4} for LAEs detected in MUSE-Wide survey at $3.3<z<6$. Plugging in the values in equation \ref{eq:virial radius}, the typical virial radius for a galaxy at $\langle z \rangle =4.12$ is $R_{vir}=41.5$ pkpc. The obtained value shows that most of the \ion{C}{iv} and \ion{Mg}{ii} systems lie beyond the virial radius of the associated galaxies with majority of them at $\sim2-4~R_\text{vir}$ of the LAEs.

\subsection{Galaxy-centric analysis: Equivalent width - Impact parameter relation}
\label{subsec:W-D}
The relation between equivalent width of the metal absorbers and the impact parameter between them and the host galaxies has been extensively studied at low redshifts $z<2$ \citep{bordoloi2011, nielsen2013, bordoloi2014, dutta2020, dutta2021, cherrey2025}. These works have shown that the absorber strength decreases with increase in impact parameter. There are not many works that focus on high redshift metal absorbers and their relation with host galaxies. \citet{bordoloi2023} studied the cool gas around galaxies at $2.3<z<6.3$ using a single sightline J0100+2802 from deep JWST/NIRCam slitless grism spectroscopy. They find galaxies associated with seven \ion{Mg}{ii} absorption systems, including 5 weak systems, within 300 kpc of the quasar line of sight. They observed that the absorber strength falls with radial separation from the galaxies. However, works such as \citet{banerjee2023, beckett2024} report little to no anti-correlation between \ion{C}{iv}/\ion{Mg}{ii} absorber strengths and impact parameter at $0.5<z<4$. 

Motivated by the results from literature, this work investigates how the absorber strength ($W$) varies as a function of impact parameter ($D$) at $2.9<z<6.7$. We focus on the two ions with most of the galaxy associations in our sample, \ion{Mg}{ii} and \ion{C}{iv}, to study the $W-D$ relation. 

\subsubsection{\ion{Mg}{ii} associated galaxies}
\label{subsubsec:MgII}

\ion{Mg}{ii} traces galaxy CGM either as outflows or inflows and therefore, should be located in the near vicinity of host galaxies \citep{zibetti2007, weiner2009,nestor2011,noterdaeme2010, bordoloi2011, Chen2010, rubin2010, bordoloi2011, bordoloi2014b, kapzak2012, nielsen2015, zabl2019, chen2025}. In this work, 14 \ion{Mg}{ii}-galaxy associations are detected within $\pm1000\text{ km s}^{-1}$ in the redshift range $3.4<z<5.1$ with a projected distance between the absorber and the associated galaxy up to $~223$ pkpc. Among them, 3 \ion{Mg}{ii} systems have multiple galaxy associations while 6 of them are isolated LAE-absorber pairs. For LAEs with no absorber counterparts, a $3\sigma$ upper limit is estimated on the absorber equivalent width centred at the LAE redshift using velocity interval of $200~\text{km s}^{-1}$ (Majority of E-XQR-30 absorbers have their 90\% velocity width ($v_{90}$) less than 200 $\text{km s}^{-1}$ \citep{xqr30catalog} and therefore, this value gives a conservatively high estimate for the equivalent width upper limits).
\begin{figure*}

\begin{subfigure}{\columnwidth}
\includegraphics[width=\columnwidth]{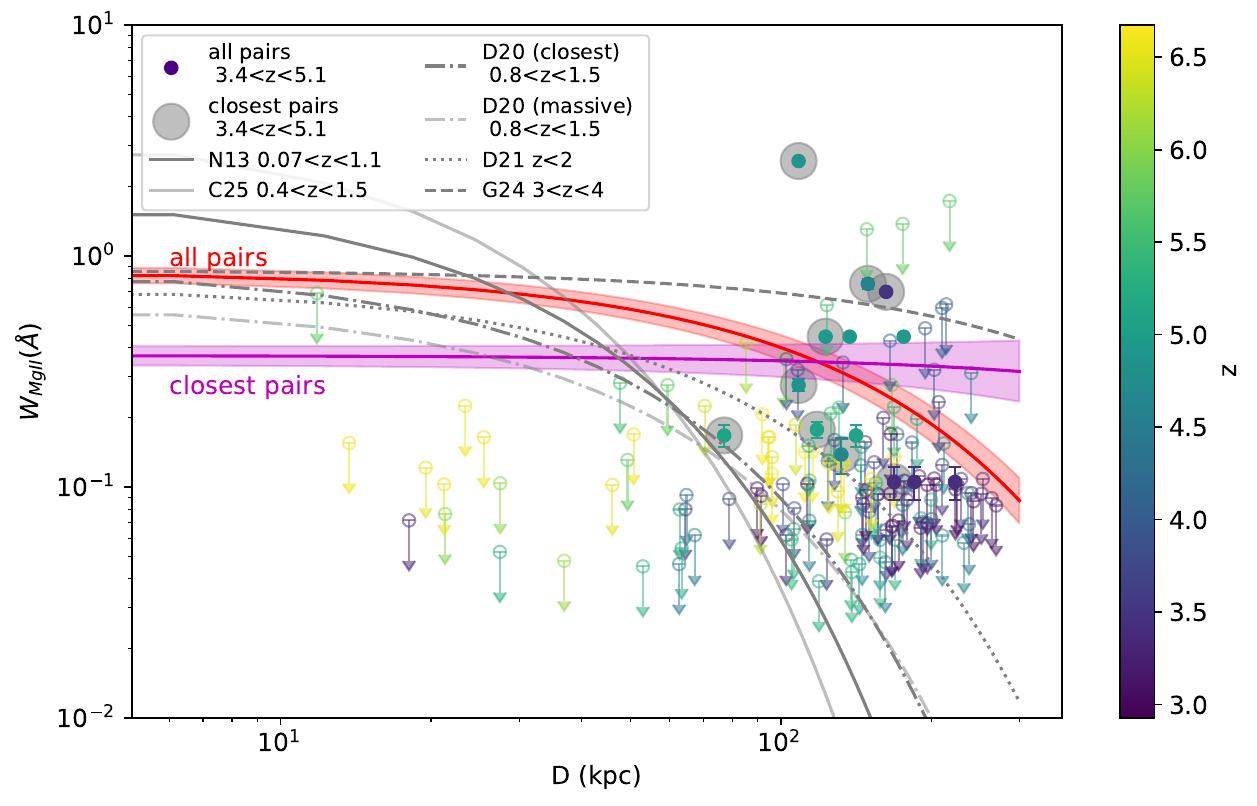} 
\end{subfigure}
\begin{subfigure}{\columnwidth}
\includegraphics[width=\columnwidth]{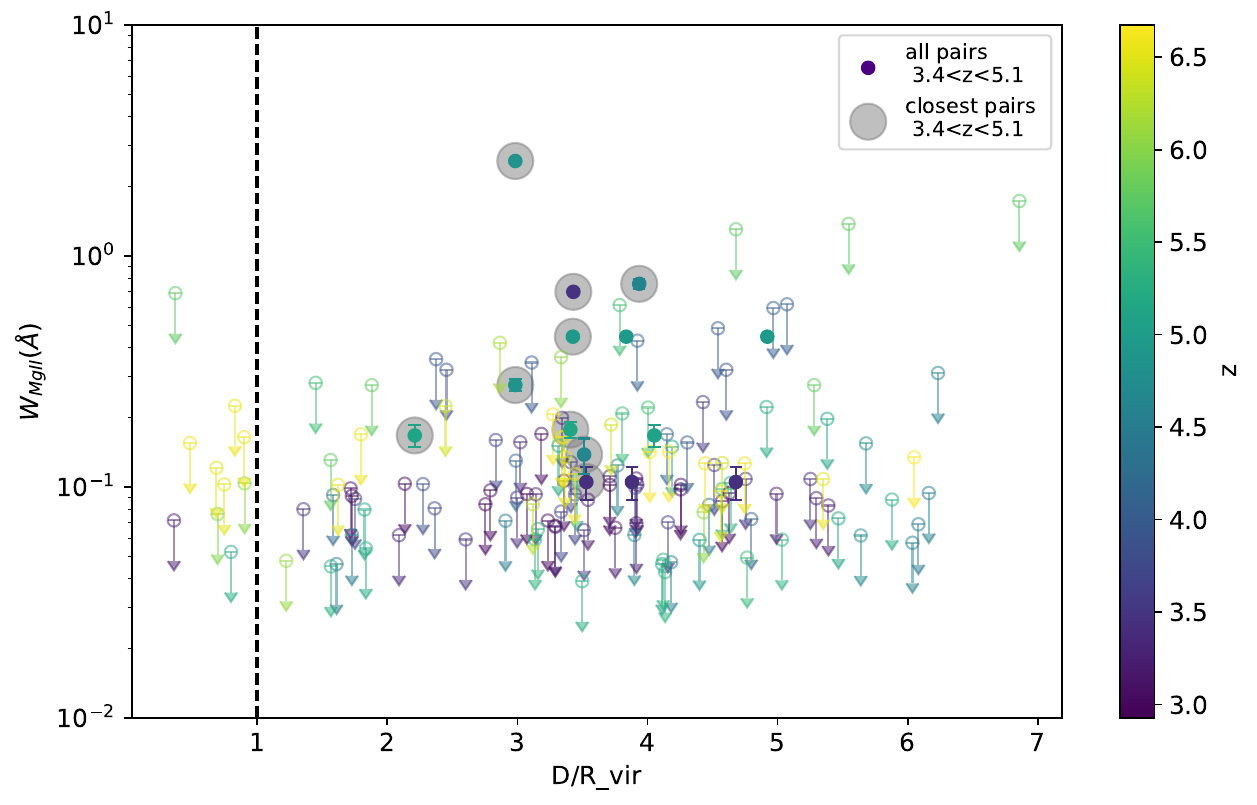}
\end{subfigure}

\caption{The equivalent width versus impact parameter for \ion{Mg}{ii} in the redshift range of $3.4<z<5.1$. The data points are colour coded based on their redshifts. The closest absorber-galaxy pairs are highlighted in grey. \textit{Left panel}: \ion{Mg}{ii} absorber strength as a function of impact parameter and are fit using a log-linear model excluding non-detections for galaxy-absorber pairs (red solid line) and for the closest galaxy-absorber pairs (magenta line). The log linear relations from various literature are shown for comparison using grey lines. A shallow decline is observed in absorber strength for \ion{Mg}{ii} towards large impact parameters compared to the low redshift literature.\textit{Right panel}: Equivalent width as a function of impact parameter normalised using the typical virial radii of the galaxies at the probed redshift range. The dashed vertical line indicates the virial radii of the LAEs. }
\label{fig:W_D_MgII}
\end{figure*}

Figure \ref{fig:W_D_MgII} shows how the absorber strength varies as a function of impact parameter (left) and as a function of impact parameter normalised by virial radii (right panel) for \ion{Mg}{ii}. The absorbers are colour coded depending on their respective redshifts.  Among the multiple galaxy associations to a single absorber, the galaxy-absorber pair with the smallest impact parameter is highlighted in grey. The non-detections of absorbers around associated galaxies are marked using upper limits which are also colour coded based on redshifts. A bimodal population of \ion{Mg}{ii} absorbers is suggested due to most non-detections reading quite weak $W$ values, whereas most detected \ion{Mg}{ii} systems have $W>0.1$\AA. Therefore, only detected measurements are used to obtain the fit. The data are fit using a log-linear model given in equation \ref{eq:log-linear fit} for all data as well as the closest pairs of absorbers and associated galaxies.

\begin{table}
    \centering
    \caption{The best-fit parameters for the log-linear fit used for the correlation between absorber strength and impact parameter (equation \ref{eq:log-linear fit}).}
    \begin{tabular}{cccc}
    \hline
      ion   & pairs & a$\times10^{-3}$ & b$\times10^{-1}$\\
      \hline
       \multirow{2}{*}{\ion{Mg}{ii}}  & all  & $-3.3_{-0.2}^{+0.2}$ & $-0.6_{-0.3}^{+0.3}$\\
         & closest & $-0.2_{-0.3}^{+0.3}$ & $-4.3_{-0.4}^{+0.4}$\\
         \hline
        \multirow{2}{*}{\ion{C}{iv}} & all & $-2.66_{-0.07}^{+0.07}$ & $-3.77_{-0.09}^{+0.09}$\\
         & closest & $-4.18_{-0.09}^{+0.09}$ & $-3.10_{-0.12}^{+0.01}$\\
         \hline
    \end{tabular}
    
    \label{tab:log-linear}
\end{table}
\begin{equation}
\label{eq:log-linear fit}
    \text{log}\ W = a\times D+b
\end{equation} The best fit parameters a and b are determined using MCMC technique with the help of \verb|emcee| Python package. They can be found in Table \ref{tab:log-linear}. For the closest absorber-galaxy pairs (red solid line), there appears to be no correlation between absorber strength and their spatial distribution around the galaxies from this dataset. When considering all pairs (magenta solid line) a mild anti correlation is observed. The right panel in Figure \ref{fig:W_D_MgII} shows that none of the \ion{Mg}{ii} absorbers (without accounting for the upper limits) are located within the virial radii of the galaxies at $3.4<z<5.1$. The only strong \ion{Mg}{ii} system with an associated galaxy in our sample is present at an impact parameter greater than twice the virial radius. 

The log-linear fit obtained in this work is compared to the results from previous works (grey lines in Figure \ref{fig:W_D_MgII}). \citet{nielsen2013} (N13) studied 182 galaxies at $0.072\le z \le1.120$ using the \ion{Mg}{ii} Absorber-Galaxy Catalog (MAG{\scriptsize II}CAT). They parametrised the strong anti-correlation between $W$ and $D$ for \ion{Mg}{ii} using a log-linear fit. Recent work by \citet{cherrey2025} (C25) reported a similar steep anti-correlation between \ion{Mg}{ii} absorber strength and impact parameter using star-forming galaxies with log $(M_*/M_\odot)>9$ from the MusE GAs FLOw and Wind (MEGAFLOW) survey at $0.4<z<1.5$. At $0.8<z<1.5$, \citet{dutta2020} (D20) detected 228 galaxies from the MUSE Analysis of Gas around Galaxies (MAGG) survey and found an anti-correlation between absorber strength and impact parameter for both cases of massive galaxies and closest galaxies associated with \ion{Mg}{ii}. Using data from the MAGG survey, \citet{dutta2021} showed that multiple galaxies are observed to be associated with single \ion{Mg}{ii} systems making it ambiguous to define a galaxy-absorber pair. They found an increase in the scatter in the W-D space which has been observed in our dataset too. On average, the absorber strength decreased with increasing distance from the closest associated galaxy for their sample. As a follow up, using \ion{Mg}{ii} data from the same survey, \citet{galbiati2024} (G24) analysed the absorbers strength against projected distance from the galaxy and observed that \ion{Mg}{ii} equivalent width drops beyond the galaxy virial radii but flattens towards large impact parameters. All the above mentioned works have used log-linear fit to describe the relation between impact parameter and equivalent width. \citet{bordoloi2023} used a power law fit instead, to indicate a steep decline in absorber strength with impact parameter using data from a broader redshift range of $2.3<z<6.3$ compared to this and other works. We attempted to fit our data using a power law function but the log-linear model proved to be a better fit. Compared to the $z<2$ studies, weak \ion{Mg}{ii} systems at high redshift tend to have larger spatial separations from the host galaxies. Our results agree with \citet{beckett2024} who has shown that there is no anti-correlation between impact parameter and equivalent width for \ion{Mg}{ii} using data from MUSE Ultra Deep Field (MUDF) at $0.5<z<3.2$ using a small sample size.

Splitting our \ion{Mg}{ii}-galaxy pairs based on absorber strength, there are 8 weak ($W<0.3$\AA), 5 medium ($0.3<W<1.0$\AA) and 1 strong ($W>1.0$\AA) \ion{Mg}{ii}-LAE associations in our sample. 
Weak \ion{Mg}{ii} absorbers are considered to be accreting or co-rotating in the outer regions of galaxy halos \citep{kapzak&churchill, kapzak2012, nielsen2015} at low redshifts. We find that the distance between an \ion{Mg}{ii} absorber and the nearest LAE - measured in virial radii - is not dependent on the absorber strength. We find weak systems closer to galaxies than strong \ion{Mg}{ii} systems. This result holds despite that the absorbers are drawn from the largest sample of weak \ion{Mg}{ii} absorbers at high redshift \citep{sebastian2024}. These absorbers showed a constant evolution in their comoving line densities ($dn/dX$) across redshift 2 to 6 which means that the mechanisms producing these weak systems should consistently replenish them \citep{sebastian2024}. The cross-matching of \ion{Mg}{ii} absorbers with LAEs has not shed any light on the constant weak \ion{Mg}{ii} $dn/dX$ across all redshifts.

 \subsubsection{\ion{C}{iv} associated galaxies}
\label{subsubsec:CIV}
\ion{C}{iv} is a popular tracer of galaxy over densities \citep{adelberger2005, diaz2021, Kashino2023, bielby2020} and the absorber strength has been shown to decline with impact parameter similar to \ion{Mg}{ii} at $z<2$ \citep{bordoloi2014, dutta2021, galbiati2023}. 27 LAEs in our sample have one or more \ion{C}{iv} in their near vicinity within $\pm1000 \text{ km s}^{-1}$ at $4.5<z<5.8$. In total, there are 34 LAE-\ion{C}{iv} associations in this sample. Among them, 12 \ion{C}{iv} systems are matched with isolated LAEs while the remaining 6 \ion{C}{iv} absorbers have multiple associated galaxies within $\sim240$ pkpc. 
\begin{figure*}

\begin{subfigure}{\columnwidth}
\includegraphics[width=\columnwidth]{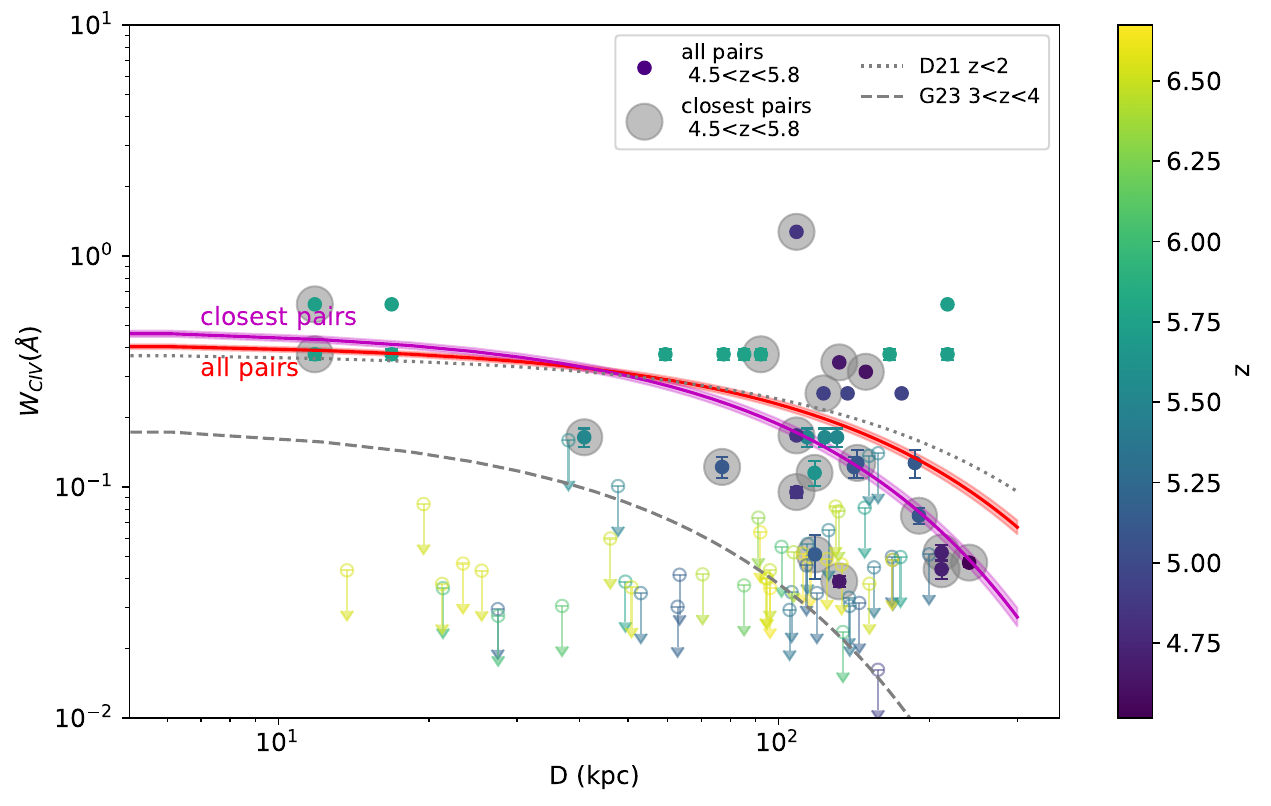} 
\caption{}
\end{subfigure}
\begin{subfigure}{\columnwidth}
\includegraphics[width=\columnwidth]{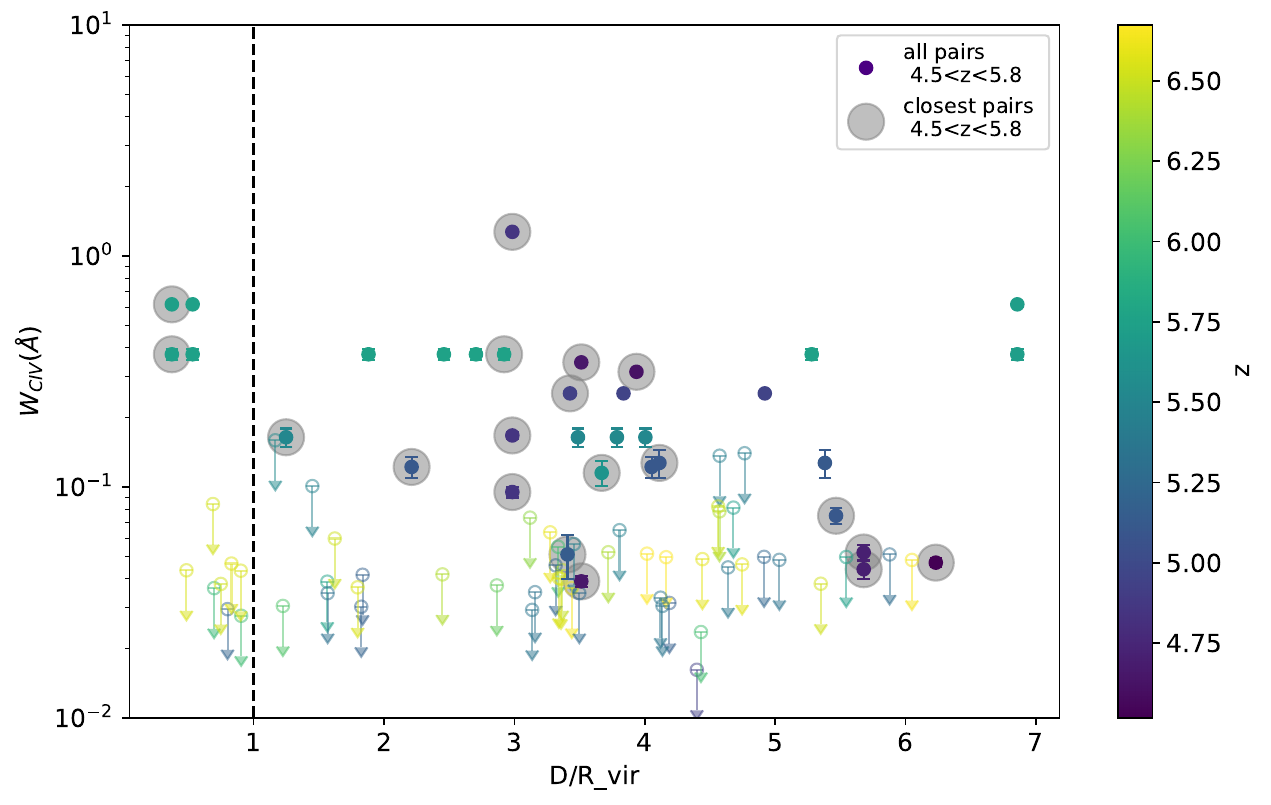}
\caption{}
\end{subfigure}

\caption{Same as Figure \ref{fig:W_D_MgII} but for \ion{C}{iv} at a redshift range of $4.5<z<5.8$. \textit{Left panel}: A flat correlation is obtained at low impact parameters ranging up to $\sim~100$ pkpc using  a log-linear fit to detected data for absorber-galaxy pairs. \textit{Right}: There are 4 \ion{C}{iv} associations within the virial radii of LAEs.}
\label{fig:W_D_CIV}
\end{figure*}

In both plots in Figure \ref{fig:W_D_CIV}, the data points are colour coded based on redshifts and the closest galaxy-absorber pairs are highlighted in grey. The galaxies for which no absorber counterparts were detected, are indicated using upper limits in absorber strength like \ion{Mg}{ii}. As shown in the left panel in Figure \ref{fig:W_D_CIV}, the log-linear fits obtained for all galaxy-absorber pairs (red curve) and the closest pairs (magenta curve) are similar except that the closest absorber-galaxy pairs have a slightly steeper slope at large impact parameters. Similar to \ion{Mg}{ii}, only detected measurements are included in the fit due to the presence of a bimodal population in W-D plane. The obtained log-linear fit is compared with literature shown using grey curves. Surprisingly, the relation obtained between $W$ and $D$ in this work aligns with the \ion{C}{iv} population at $z<2$ (grey dotted line) studied by \citet{dutta2021} using the Quasar Sightline and Galaxy Evolution (QSAGE) survey. Furthermore, there are other low and high redshift works in the literature showing a lack of dependence for \ion{C}{iv} absorber strength on impact parameter \citep{muzahid2021, banerjee2023, beckett2024} either due to lack of physical connection between these absorbers and galaxies or due to the presence of fainter galaxies closer to the absorbing gas. Comparing our results with \citet{galbiati2023} (grey dashed line) in which \ion{C}{iv} and \ion{Si}{iv} absorbing gas linked to 143 LAEs at $3<z<4$ were analysed using MAGG survey, the trend line is largely flatter with a higher normalisation than their model and a larger extent in spatial distribution is observed for \ion{C}{iv} in our work. The right panel in Figure \ref{fig:W_D_CIV} shows the equivalent width of \ion{C}{iv} as a function of impact parameter normalised using virial radii calculated using equation \ref{eq:virial radius}. There are four \ion{C}{iv}-LAE pairs that are located within the virial radii at $z\sim5.7$.

\citet{dutta2021} found that \ion{C}{iv} absorber strength declines shallower than \ion{Mg}{ii} as a function of projected distance from the host galaxies at $z<2$ proposing that the high ionised gas has more spatial extent than low ionised gas. Low redshift theoretical and observational works have shown that cool gas tends to be located closer to galaxies compared to the ionised phase where it extends to several times larger than the virial radius of the galaxy \citep{weng2024, kacprzak2025}. On the contrary, this work creates a picture of similar radial distribution of both ions independent of impact parameter although these ions trace different gas densities. However, the more ionised \ion{C}{iv} ($\sim65\%$) has higher fraction of multiple associated galaxies compared to \ion{Mg}{ii} ($\sim57\%$) at $z>3$ similar to what has been observed in \citet{muzahid2021, banerjee2023, galbiati2024}. Measuring the covering fraction of these absorbers will help us to understand the presence of galaxy groups around these ions. However, we do not have enough sample size to perform covering fraction analyses as a function of impact parameter and their evolution across redshift. For both ions, no associated galaxies are detected towards $z>5.8$, indicating that these absorbers might be connected to galaxies fainter than the detection threshold of this work at higher redshifts. Additionally, the non-detections of absorbers around galaxies at impact parameters less than $R_\text{vir}$ for both \ion{Mg}{ii} and \ion{C}{iv} suggests these absorbers may not be physically bound to the associated galaxies but rather indicate the larger scale environment. Nevertheless, outflows from galaxies at these redshifts can reach to large velocities enriching the gas at large impact parameters even beyond the virial radii as reported in observations and simulations \citep{booth2012, keating2016, carniani2024}.

\subsection{Stellar masses of the associated galaxies}
\label{subsec: stellar masses}
The stellar masses of the host galaxies can be calculated by converting the rate of photon counts to stellar masses in regions corresponding to the LAE coordinates detected from MUSE datacubes (see Figure \ref{fig:keck images}). The redshift range for the detection of LAEs via NIRC2 imaging is confined to $2.9 < z < 4$. Beyond this redshift interval, the sensitivity of the instrument diminishes, rendering it less capable of detecting galaxies with luminosities of $0.2L_*$. This limitation arises because the majority of host galaxies exhibiting \ion{Mg}{ii} are anticipated to possess luminosities exceeding this threshold, as documented by \citep{nielsen2013}. Due to the spectral coverage restriction, only \ion{Mg}{ii} associated galaxies can be observed because \ion{C}{iv} associated galaxies are detected at $z>4$ in this work. No LAEs were detected in the NIRC2 images from either H and K$_\text{p}$ filters and therefore, upper limits for stellar masses are estimated separately for each filter. 

To calculate the upper limits, an aperture size with radius of 0.5" is used for each LAE. The photon counts per second in each aperture is added in quadrature to deal with the negative counts within the aperture. The obtained sum is then converted into apparent magnitudes using the equation below \begin{equation}
    m_{\text{filter, AB}}=ZPt-2.5\text{log}_{10}\bigg(\sqrt{\Sigma~\text{(counts/sec)}^2}\bigg)
\end{equation} where $ZPt$ is the zero-point which is equal to 24.8 for H filter and 24.2 for K$_\text{p}$ filter. The apparent magnitudes are then converted into absolute magnitudes using the distance modulus formula given as \begin{equation}
    M_{\text{filter, AB}} = m_{\text{filter, AB}}-5\text{log}_{10}(d)+5
\end{equation} where d is the distance in parsecs calculated from the redshifts of the LAEs. The absolute magnitudes of the LAEs are found to be within the range of 25.4 and 25.7 for H and 24.2 and 24.9 for K$_\text{p}$. The absolute magnitudes are converted to luminosities in solar units using the formula, \begin{equation}
    L = 10^{-0.4(M_{\text{filter, AB}}-M_{\text{sun}})}
\end{equation} where $M_{\text{sun}}$ is the solar magnitude corresponding to 3.327 for H and 3.356 for K$_\text{p}$ filters \citep{Into2013NewDust}. These luminosities can be converted to stellar masses using stellar mass-to-light ratios from \citet{Into2013NewDust} given as \begin{equation}
\label{eq:stellar mass to light ratio}
    M_* = 10^{(s\times \text{colour}+z)}\times L
\end{equation} where $s$ and $z$ are coefficients from fitting colour-$M_*/L$ relation (CMLR). In rest frame, the (H-K$_\text{p}$) colour transform to (g-r) in the rest frame, so the coefficients corresponding to the latter from \citet{Into2013NewDust} is used where $s=1.774$ and $z=-0.783$. The (g-r) colour for star forming galaxies is observed to be $\sim0.5$ \citep{galaxyzoo2010}. The values are substituted in the above equation \ref{eq:stellar mass to light ratio} to calculate upper limits for stellar masses. For the H filter, the stellar mass upper limits of the associated galaxies are calculated as $\text{log}M_*/M_\odot<10.2$ and for the K$_\text{p}$ filter, the upper limit estimate is $\text{log}M_*/M_\odot<10.7$. These upper limits indicate that the probed LAEs are faint, low mass galaxies rendering them invisible in the NIRC2 images for the observed exposure time. The estimated upper limits in this work are in agreement with the measured low stellar mass values at $z>3$ from \citep{ono2010, liu2024, iani2024}. Another assumption based on the stellar mass upper limits is that these LAEs might have sub-solar metallicity based on stellar mass metallicity relations of star forming galaxies at $z>3$ \citep{nagamine2010, Laskar_2011, sommariva2012}. 

\section{Discussion}
\label{sec:discussion}

\subsection{Flat correlation between absorber strength and impact parameter}
\label{subsec:flat W-D}
We observed a weak anti-correlation between absorber equivalent width and impact parameter for both \ion{Mg}{ii} and \ion{C}{iv} at $3.4<z<5.8$ 
We do not find any strong absorber - LAE pairs at small impact parameters at high redshifts (except for two relatively strong systems for CIV-LAE pairs). Our findings contradicts the popular concept established at lower redshifts of strong absorbers residing closer to the galaxy disks while weak absorbers residing in the outer regions of the CGM \citep{nielsen2013, dutta2020, bordoloi2023}. 

The comparison of different literature results on the $W$-$D$ correlation is not straightforward as there are differences in the way galaxies are selected; some analyses are galaxy-centric \citep{bordoloi2014, dutta2021} while others are absorber-centric \citep{bordoloi2023, cherrey2025} or both \citep{nielsen2013, dutta2020, galbiati2023}. There are also dissimilarities in the ways how galaxies are detected and how the association is defined. While $z<2$ works used \ion{O}{ii} emission lines to identify galaxies \citep{dutta2020, dutta2021}, higher redshift works used Lyman $\alpha$ emission \citep{galbiati2023, galbiati2024} and observed frame near-infrared lines such as \ion{O}{iii} \citep{bordoloi2023}. These detected emission lines are sensitive to the galaxy properties and their surrounding environment. Moreover, the definitions of galaxy-absorber associations vary across literature with certain works using $\pm500~\text{km s}^{-1}$ \citep{dutta2020, dutta2021, beckett2024, galbiati2023, galbiati2024, muzahid2021} while others use $\pm1000~\text{km s}^{-1}$ \citep{mackenzie2019, bielby2020, Kashino2023, Zou2024AVLT}. The flatter correlations obtained in this work, similar to \citet{muzahid2021, banerjee2023, beckett2024}, for both ions might be caused as a result of missing a large sample of galaxy-absorber pairs by limiting ourselves to LAEs only. Future high-redshift surveys combining multi-wavelength probes for galaxy detections \citep[for e.g.,][]{bordoloi2023} (but with more sightlines) can provide deeper insights into the inconsistent trends among various literature. 

Another notable feature of this work is that the steepness of the anti-correlation flattens towards higher redshifts for Mg {\scriptsize II} as can be seen in the left panel of Figure \ref{fig:W_D_MgII}. Although our trendline at high redshift for C {\scriptsize IV} (see left panel in \ref{fig:W_D_CIV}) falls in line with low redshift trend from \citet{dutta2021}, the flat correlation at low redshift can be accounted to the lack of strong upper limits in their data. Given that the field of view for detecting galaxies would be larger at low redshifts, galaxy-absorber pairs can be detected to large distances which might also cause the flat trend line at low redshift \citep{galbiati2023}. 

No \ion{Mg}{ii} absorbers are detected within the virial radii of 10 LAEs whose galaxy halos were intersected by the lines of sight as shown in the right panel of Figure \ref{fig:W_D_MgII}. On the contrary, 4 out of 12 LAEs have \ion{C}{iv} within the virial radii of their galaxy halos. Comparing the presence of absorbers within the virial radii of the associated galaxies, we expect that the covering fraction of \ion{Mg}{ii} systems at high redshift is very low as these systems trace small cool clumps of gas lowering their chances of being intersected the quasar sightline. High redshift \ion{C}{iv} systems might be tracing higher temperature gas clouds with larger covering fraction increasing their chances of being captured in the quasar spectra. A similar finding has been reported by \citet{galbiati2024} where \ion{C}{iv} has a higher covering fraction compared to \ion{Mg}{ii}.

The lack of galaxies detected close to strong absorbers also contribute to the observed flat $W-D$ correlation. Among the 5 strong \ion{Mg}{ii} absorbers detected in three quasar sightlines, only one has an associated galaxy detected in this work. Although star-forming galaxies are expected around strong \ion{Mg}{ii} absorbers \citep{chen, sebastian2024}, the lack of galaxy associations for \ion{Mg}{ii} at $W>1.0$\AA~ is possibly due to the host galaxies remaining undetected due to three reasons. The first reason is that strong \ion{Mg}{ii} traces optically thick gas at high redshifts 
\citep{cooper, sebastian2024}. Therefore, the Lyman $\alpha$ line from the host galaxy is completely attenuated by the dense metal-enriched neutral gas surrounding the galaxies, rendering them undetectable \citep{keating2020}. The only strong \ion{Mg}{ii} system with an associated galaxy is a \ion{C}{iv} dominated system, which may have created a more ionised environment for the Lyman $\alpha$ photons to escape. The second reason is that these galaxies can be dusty and thereby not detected in this work. At low redshift, \citet{menard2012} found that strong \ion{Mg}{ii} absorbers carry a significant amount of dust expelled by galaxies. The high detection of \ion{C}{ii} emitters detected around enriched neutral regions \citep{Neeleman2017, Neeleman2019}  provides hints of a significant population of dust obscured galaxies located in dense regions of the Universe. Dedicated research on these dust attenuated galaxies should be performed to understand more about the host galaxies around strong \ion{Mg}{ii} using Atacama Large Millimeter/submillimeter Array (ALMA) and JWST. The third reason could be that the associated galaxy is fainter than the detection limit of this work - $\text{log}~F[\text{ergs}^{-1}\text{cm}^{-2}]<-17.79$. 

\begin{figure}
    \centering
    \includegraphics[width=\columnwidth]{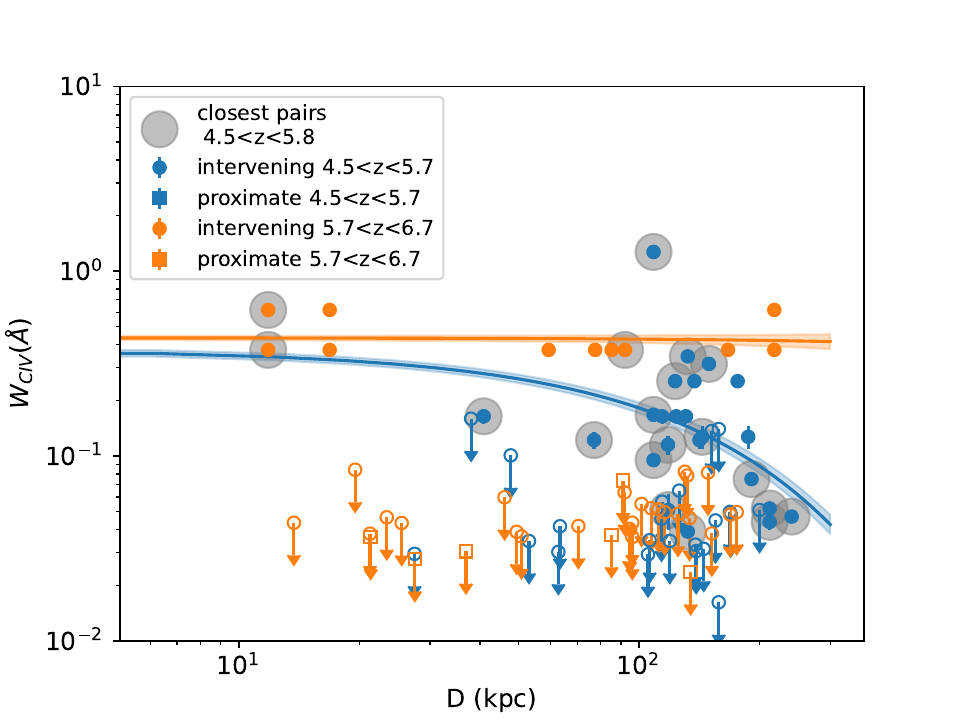}
    \caption{Correlation between \ion{C}{iv} absorber strength ($W$) and impact parameter ($D$) after splitting the sample at $z=5.7$. The detected $z>5.7$ absorber-galaxy pairs are denoted using orange solid markers while lower redshift detections are indicated using blue filled data points. The upper limits are represented using empty markers with downward arrows and colour-coded based on their redshifts. The proximate absorber-galaxy pair are marked using squares and the intervening absorber-LAE pairs using the circles. The closest absorber-galaxy pairs are highlighted in grey. The high redshift sample show no correlation between absorber strength and impact parameter while the low redshift data points indicate a weak anti-correlation.}
    \label{fig:W_D_CIV_z}
\end{figure}

Recent works report that high redshift absorbers are dominated by low ionisation species, characterised by lower column density ratios of \ion{C}{iv} to \ion{C}{ii} at $z>5.7$ \citep{cooper, Rowlands2026E-XQR-30:Z2}. To assess the impact of high-redshift \ion{C}{iv}-LAE associations, we split our sample at $z=5.7$ and fit our LAE-absorber pairs using a log-linear model. We find that $z>5.7$ absorbers show no correlation between their strength and impact parameter while $z<5.7$ show an anti-correlation between absorber equivalent width and impact parameter as shown in Figure \ref{fig:W_D_CIV_z}. To further understand if the high redshift \ion{C}{iv} absorbers that produced a flat relation are affected by quasar proximity, the proximate \ion{C}{iv} systems (squares) that are within $10000~\text{km s}^{-1}$ from the quasar redshift are distinguished from the non-proximate absorbers (circles). But none of the detected high redshift \ion{C}{iv} absorbers with galaxy associations are proximate.

It is observed that most of the \ion{Mg}{ii} and \ion{C}{iv} absorbers are located at 100-150 pkpc from the associated galaxies causing a flatter anti-correlation between $W$ and $D$ at high redshift. It is surprising that these absorbers are located beyond the virial radii of the associated galaxies. The presence of absorbers far away from their potential host galaxies indicates the presence of fainter galaxies closer to these gas clouds that are not detected in this work. Gas dynamical simulations performed on galactic outflows from intermediate-redshift dwarf galaxies at $z=2$ shows that the weak \ion{Mg}{ii} systems are produced by these undetected satellite galaxies \citep{Fujita2020}. Similarly, theoretical model from \citet{garcia2017} predicts that galaxies observed to be separated from the absorbers at large distances might not be physically connected. Moreover, \citet{Doughty2023The5} has suggested a communal enrichment of galaxy environments by low-mass fainter galaxies rather than a single contributor to account for the presence of absorbers beyond the virial radii of the detected potential host galaxies. We examined the \ion{O}{iii} catalogue from \citet{Kashino2025} for galaxies closer to our lines of sight than the LAEs in this work. Although, 54 \ion{O}{iii} emitters are found to have \ion{Mg}{ii} and/or \ion{C}{iv} absorbers within $\pm1000~\text{km s}^{-1}$, none of them are located closer than the LAEs detected in this work.

Investigating the physical reasons for observing a weak $W$-$D$ anti-correlation at z>3.4, the velocities of galaxy outflows at these redshifts have also to be considered. 
If the mean outflow velocity for typical host galaxies is $200~\text{km s}^{-1}$ \citep{furlanetto2003} starting at $z=10$, it is expected for the absorbers to reach up to a distance of $\sim100$ pkpc at $z\approx5.8$ and to a distance of $\sim300$ kpc at $z\approx3.4$. However, this holds true only if the galaxy outflows are continuous throughout the cosmic period between $z=10$ and $z=3.4$ which is unlikely due to their episodic nature. Recent work by \citet{carniani2024} reported a median velocity of $350~\text{km s}^{-1}$ for galaxies with halo masses between $10^7~M_\odot$ and $10^8~M_\odot$ at $z>4$ using the JWST Advanced Deep Extragalactic Survey (JADES). Given the typical halo mass of LAEs observed at $3.3<z<6$ \citep{lsdcat4}, the outflow velocity could be $\sim350~\text{km s}^{-1}$ depending on the halo mass \citep{mitchell2020}, depositing metals further outward or inward of the CGM. 
For \ion{Mg}{ii}, a wind speed of $200~\text{km s}^{-1}$ starting at $z=10$ can carry the gas to the observed impact parameters while for \ion{C}{iv}, an outflow velocity of $350~\text{km s}^{-1}$ starting at $z=9$ is required. However, additional work, both in observations and simulations, has to be done to provide robust constraints on the outflow velocities of galaxies at these redshifts.

\subsection{Is there an LAE overdensity around metal absorbers?}
\label{subsec:galaxy overdensity}

Previous works on galaxy-absorber connections find that LAEs cluster around strong absorption systems and lie preferentially near metal line absorbers than random regions \citep{Fumagalli2024ThePerspective}. Observational \citep{diaz2015, bielby2020, Lofthouse2022MUSEGalaxies, galbiati2023, galbiati2024, Zou2024AVLT, Banerjee2025} and theoretical works \citep{finlator2020, Doughty2023The5} have reported an overdensity of galaxies around metal absorbers, particularly \ion{C}{iv} and \ion{Mg}{ii}. 

We measure the overdensity of LAEs around metal absorbers by searching for LAEs detected within a velocity window of $\pm1000~\text{km s}^{-1}$ at random redshifts covering Mg {\scriptsize II} ($3.1<z<6.1$) and C {\scriptsize IV} ($4.5<z<6.4$) associated galaxy redshift ranges. The number of random redshift windows is determined by matching the total number of \ion{C}{iv} or \ion{Mg}{ii} systems within the redshift range along each line of sight. The redshift windows where associated galaxies are detected are omitted when probing for LAEs in random regions. We match the number of random windows generated for each line of sight with the number of absorbers detected in the E-XQR-30 catalogue for that quasar at $z>2.9$. The overdensity ratio is then calculated by dividing the number of absorber-LAE associations around the respective ions by the number of LAEs detected in random redshift windows. After repeating this process 100 times per cube, we find the LAE overdensity ratios for \ion{C}{iv} and \ion{Mg}{ii} as 1.7 and 1.9 respectively. The values used in the overdensity calculations can be found in Table \ref{tab:overdensity} corresponding to a mild increase in LAE numbers around the absorbers. We defer to the complete REQUIEM survey for robust constraints on the overdensity ratios of LAEs around absorbers due to large field-to-field variations of our sample. 

\begin{figure*}
    \centering
    \includegraphics[width=\linewidth]{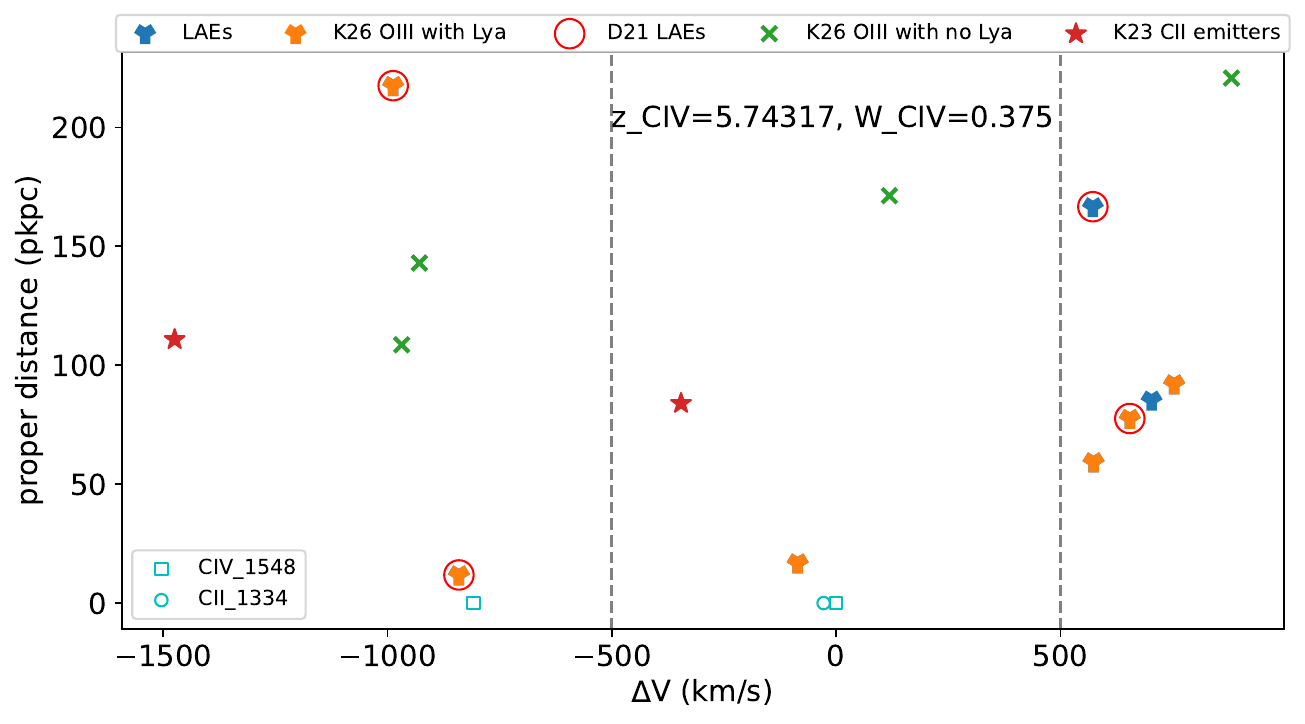}
    \caption{The \ion{C}{iv} absorption system in J1030 at $z=5.74$ with an LAE overdensity ratio of $\approx10$. The redshift is centred around \ion{C}{iv}. The strongest absorber in the system, \ion{C}{ii}, is also shown in the plot. There is another \ion{C}{iv} system at $z=5.73$ within $\sim800~\text{km s}^{-1}$ from the central \ion{C}{iv} system. The LAEs that are already reported in \citet{diaz2021} are encircled in red. The associated LAEs are marked using blue tridents where those detected as \ion{O}{iii} emitters in \citet{Kashino2025} are marked in orange. The \ion{O}{iii} emitters with no detected Lyman $\alpha$ emission are indicated using the green crosses. The red stars show the two \ion{C}{ii} emitters reported in \citet{Kashino2023}. The vertical dashed lines indicate $\pm500~\text{km s}^{-1}$ velocity separation.} 
    \label{fig:J1030 LAE overdensity}
\end{figure*}

Notably, there is an interesting \ion{C}{iv} absorption system at $z=5.74$ in J1030 with 8 LAEs within $\pm1000~\text{km s}^{-1}$ as shown in Figure \ref{fig:J1030 LAE overdensity}. The system is a multiphase gas cloud with low and high ionisation absorbers among which \ion{C}{ii} is the strongest absorber \citep{xqr30catalog}. There is another \ion{C}{iv} system located at $\sim800~\text{km s}^{-1}$ from the central \ion{C}{iv} system. Restricting the above-mentioned method to J1030 for calculating LAE overdensities around absorbers, we find an overdensity ratio of $\approx10$ around this system compared to the mean of random LAEs in 39 redshift windows. The strong association of this system with LAEs was first noted in \citet{diaz2021} and some of these LAEs are also detected as \ion{O}{iii} emitters \citep{Kashino2025}. There are other \ion{O}{iii} emitters from \citet{Kashino2025} with no detected Lyman $\alpha$ emission. Two \ion{C}{ii} emitters reported in \citet{Kashino2023} are associated with these \ion{C}{iv} systems. Overall, there are 13 galaxies with a variety of types sitting within a significant LAE overdensity associated with two \ion{C}{iv} systems. The closest galaxy to each absorber also happens to be an LAE. These two are the closest associations found across the three lines of sight in this study.
These galaxies observed in this overdense region located within a redshift window of $\Delta z=0.04$ communally enrich its surroundings as evident from the two multi-phase \ion{C}{iv} systems. Further observations are required to study the properties of this galaxy overdensity members to understand its stellar populations and other features in detail.

\begin{table*}
    \centering
    \caption{The overdensity ratio details for LAEs around metal absorbers. Random LAE numbers are from the sample of redshift windows of $\text{d}v=\pm1000~\text{km s}^{-1}$ corresponding to the number of absorbers with associated galaxies across the three lines of sight.}
    \begin{tabular}{cccccc}
    \hline
     Ion     & Redshift & No. of  & No. of LAE- & $\langle \text{random LAE counts}\rangle$ & Overdensity ratio = $\frac{\text{\#LAE-abs associations}}{\text{\#random LAEs}}$\\
     & range& absorption systems & absorber associations & & \\
    \hline
      \ion{C}{iv} & $4.5<z<5.8$   & 39 & 34 & 19.7 & 1.7\\
      \ion{Mg}{ii} & $3.4<z<5.1$   & 22 & 14 & 7.5 & 1.9\\
      \hline
    \end{tabular}
    
    \label{tab:overdensity}
\end{table*}

\section{Conclusions}

We performed a blind search for LAEs in three $z\sim6$ deep quasar fields that were observed with VLT/MUSE with exposure time between 6.4 and 10.6 hours, selected from the MARQUIS survey (Meyer et al. in prep). The absorption systems in the three quasar sightlines are obtained from the E-XQR-30 metal absorber catalogue \citep{xqr30catalog}. A total of 156 LAEs are detected across $2.9<z<6.7$ with flux ranging from log $F[\text{ergs}^{-1}\text{cm}^{-2}]=-17.8$ to -16.2 using the combination of LSDCat \citep{LSDCat1.0, LSDCat2.0}, QtClassify\citep{qtclassify} and PyPlatefit \citep{pyplatefit}. The systemic redshifts of these LAEs are computed using the correlations from \citet{Verhamme2018}. Among these, 31 galaxies have \ion{Mg}{ii} or \ion{C}{iv} or both within a velocity window of $\pm1000~\text{km s}$ at a transverse distance of $<250~\text{pkpc}$. The LAE sample completeness is assessed by creating mock cubes with fake sources in them. The sample has 50\% completeness limit at log $L [\text{ergs}^{-1}]>41.7$. We analysed the correlation between absorber equivalent width and impact parameter for \ion{C}{iv} and \ion{Mg}{ii} systems associated with LAEs. Our major findings are outlined below.
\begin{itemize}
    \item We observed a flatter anti-correlation between the absorber strength and impact parameter for \ion{Mg}{ii} associated galaxies at $3.4<z<5.1$ and \ion{C}{iv} associated galaxies at $4.5<z<5.8$ compared to the low redshift samples. This can be attributed to the non-detections of fainter or dust-obscured galaxies that are closer to these absorbers or to the high velocity galaxy outflows \citep{carniani2024}.
    \item Most of the absorbers are located at $2-3R_\text{vir}$ of the LAEs. No absorber-galaxy pairs are found within the typical virial radii of the LAEs for \ion{Mg}{ii} while for \ion{C}{iv}, there are 4 absorber-galaxy pairs within the virial radii. This indicates that \ion{Mg}{ii} absorbers have lower covering fraction than \ion{C}{iv} in the early Universe.
    \item Using the non-detections of associated LAEs in Keck/NIRC2 images, upper limits for stellar masses are computed that ranges between $\text{log}M_*/M_\odot<10.7$ indicating that these are low mass, faint sources. 
    \item Overall, we find mild overdensity ratios of 1.7 and 1.9 for LAEs around \ion{C}{iv} and \ion{Mg}{ii} respectively. An exception is noted in a \ion{C}{iv} system at $z=5.7$ in J1030, that hosts an LAE overdensity of $\approx10$ compared to the random regions of the Universe.
    
\end{itemize}
This work will serve as a preparatory study for the upcoming REQUIEM survey where 31 quasars in the southern sky with redshifts across $5.77<z<6.62$ are observed using MUSE on VLT. The pipeline developed here for the detection, classification and measurements of emission lines can be employed for this LP. The larger and better sample of galaxies from REQUIEM will enable us to extend the sample of absorber-galaxy associations to obtain robust results on their environment at high redshift.

\section*{Acknowledgements}
The authors acknowledge the anonymous referee for their valuable feedback. 
The authors thank the E-XQR-30 collaboration for observing and reducing quasar spectroscopic data used in this work. Based on observations collected at the European Organisation for Astronomical
Research in the Southern Hemisphere under ESO Programme IDs: 0103.A-0140(A), 106.215A.001 and 095.A-0714(A).

The authors thank Glenn Kacprzak and Themiya Nanayakkara for the valuable discussions. AMS acknowledges the support from Edmund Christian Herenz and Hector Salas Olave in helping to resolve the issues related to LSDCat and QtClassify. The authors also appreciate the efforts of Anna-Christina Eilers and Daichi Kashino for providing the required literature data for cross-matching. 

This work used NASA's Astrophysics Data System, \textsc{matplotlib} \citep{Hunter:2007}, \textsc{numpy} \citep{harris2020array}, \textsc{scipy} \citep{2020SciPy-NMeth}, \textsc{pandas} \citep{reback2020pandas}, \textsc{regions} \citep{larry_bradley_2022_6374572} and \textsc{photutils} \citep{larry_bradley_2025_14889440}. A significant usage of SAOImageds9 \citep{saoimage} and QFitsView \citep{qfitsview} has enabled the successful completion of this work. 

\section*{Data Availability}

The E-XQR-30 metal absorber catalogue used in this project can be downloaded at \href{https://github.com/XQR-30/Metal-catalogue/tree/main/AbsorptionPathTool}{https://github.com/XQR-30/Metal-catalogue/tree/main/AbsorptionPathTool}. The electronic version of the catalogue of detected LAEs in this work is available at \href{https://doi.org/10.5281/zenodo.20279200}{https://doi.org/10.5281/zenodo.20279200}.



\bibliographystyle{mnras}
\bibliography{example} 




\appendix

\section{LAE detection and classification}
\label{sec:LAE detction flowchart}

\begin{figure*}
  \centering
  \begin{subfigure}[b]{0.45\textwidth}
    \centering
    \includegraphics[width=\textwidth]{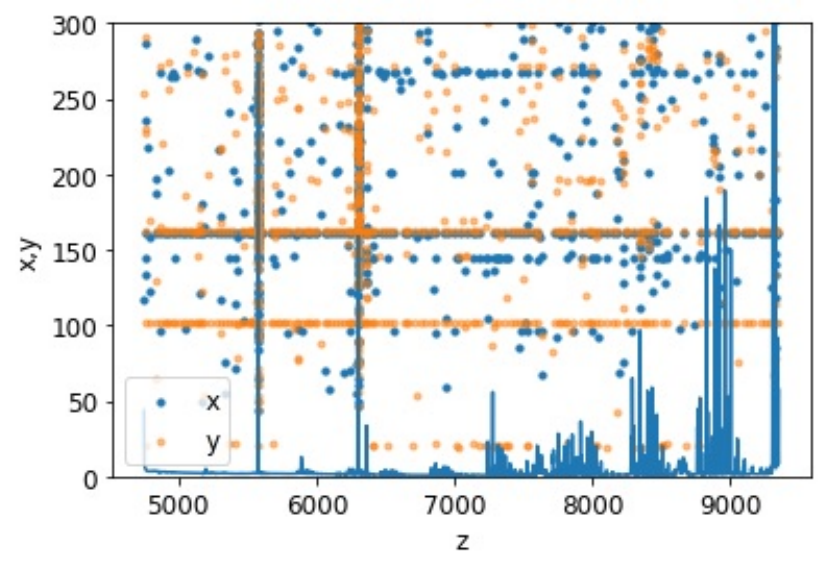}
    \caption{}
    \label{fig:excess detections}
  \end{subfigure}
  \hfill
  \begin{subfigure}[b]{0.45\textwidth}
    \centering
    \includegraphics[width=\textwidth]{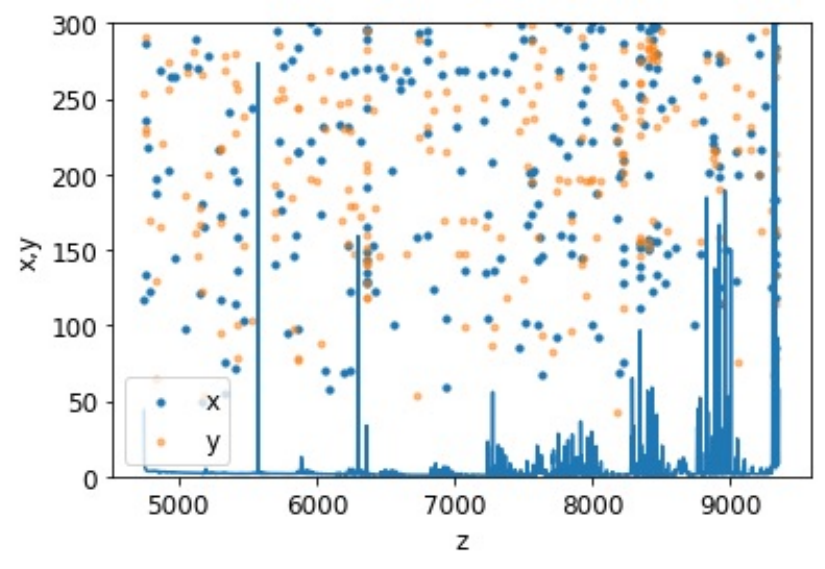}
    \caption{}
    \label{fig:no excess detections}
  \end{subfigure}
  \caption{The left figure shows the excess detections of the emission lines in the datacube J1030+0524 along certain x and y pixel coordinates and spectral channels. The right figure displays the detections after removing the excess detections in an iterative manner for the same datacube. In both figures, the blue points represent the detections along the x direction and the orange points indicate the detections in the y direction. The noise spectrum for J1030 is shown at the bottom for both figures to demonstrate that the excess along certain spectral layers correspond to the skylines.}
  \label{fig:excess and no excess detections}
\end{figure*}

On careful examination of the intermediate catalogue produced by LSDCat using the median filtered cubes, it has been observed that even the ZAP reduced cubes produced an excess detections along certain pixel values and wavelength channels, especially along the skylines (see Figure \ref{fig:excess detections}). In order to eliminate these false detections, the wavelength regions and the coordinates with excess detections were masked and then passed through LSDCat again. This process is repeated until the striped clustering of detections across different wavelength channels and spatial coordinates was reduced (see Figure \ref{fig:no excess detections}).

The difference in LSDCat outputs between masked and unmasked cubes can be seen in Figure \ref{fig:excess and no excess detections}.

\begin{figure*}
    \centering
    \includegraphics[width=1\textwidth]{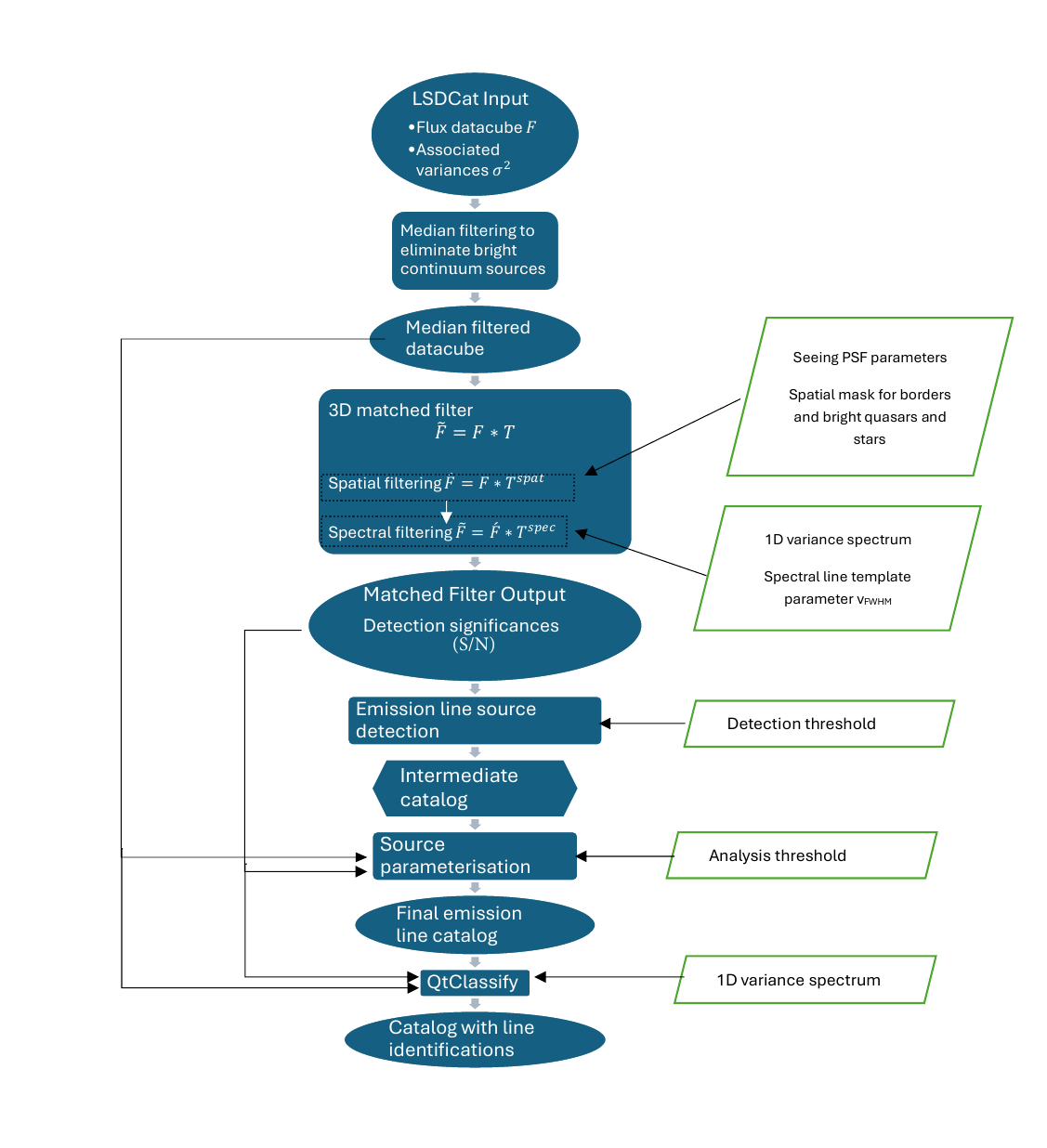}
    \caption{The flowchart showing the steps followed for detecting and identifying the emission lines from the wide-field IFU datacube.}
    \label{fig:LAE workflow}
\end{figure*}

The different routines used for detecting emission lines using LSDCat and classifying the lines using QtClassify is given as a flowchart in Figure \ref{fig:LAE workflow}.

\section{Measuring line flux using PyPlatefit}
\label{sec:PyPlatefit flowchart}
\begin{figure*}
    \centering
    \includegraphics[width=0.8\textwidth]{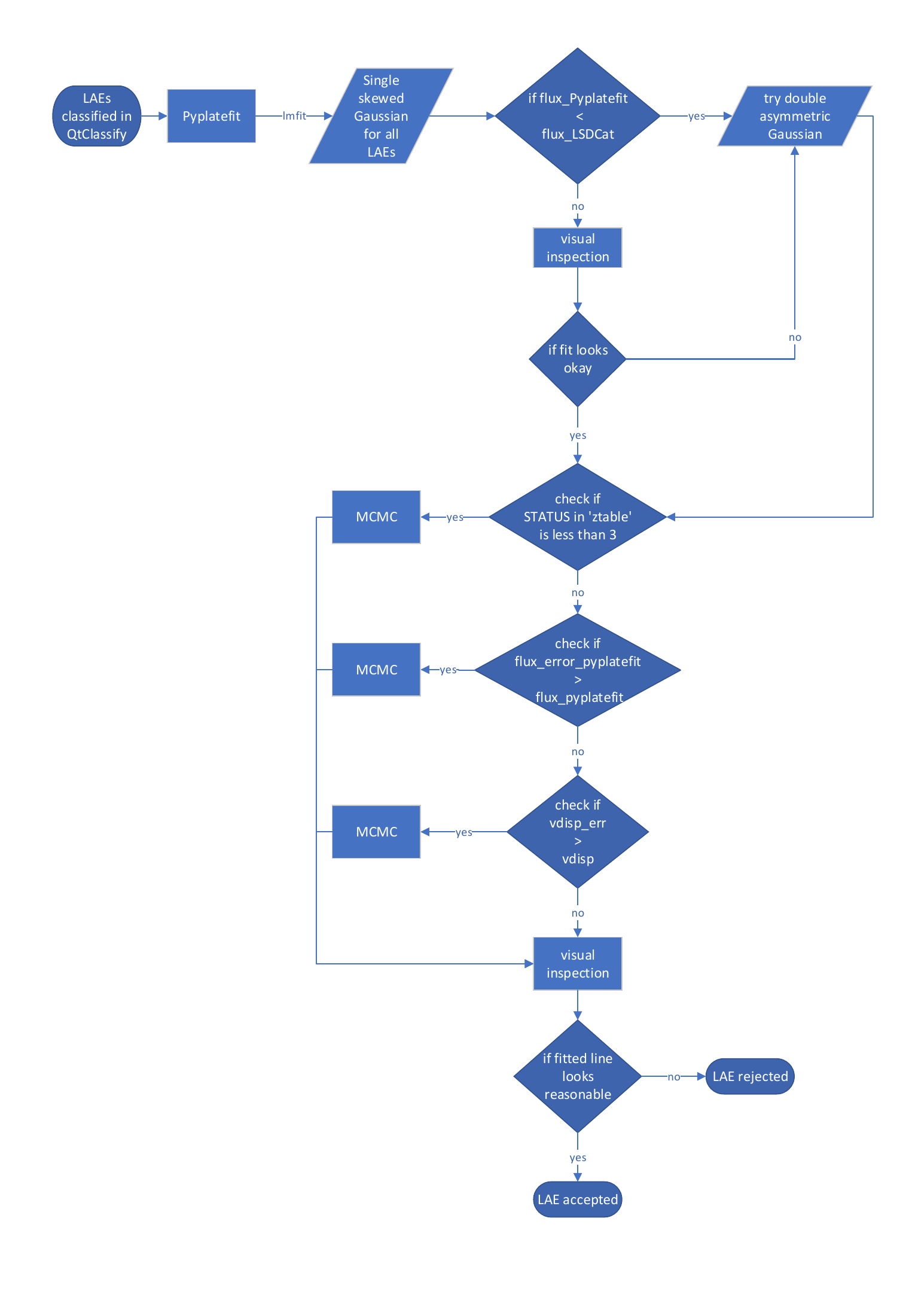}
    \caption{A flowchart showing the steps followed in fitting the Lyman $\alpha$ emission lines in pyPlatefit.}
    \label{fig:pyplatefit}
\end{figure*}

A flowchart explaining the various steps involved in measuring the flux of the LAEs is shown in Figure \ref{fig:pyplatefit}.

\section{LAE Catalogue}
\label{sec:catalogue}
The catalogue consisting of the detected LAEs with their LSDCat assigned ID, sky coordinates, emission line ($z_{\text{Ly}\alpha}$) and systemic ($z_\text{systemic}$) redshifts, flux and associated errors, flags for doublet profile and cross-matches with literature \citep{diaz2021, Kashino2025} can be found in Table \ref{tab:LAE detections}. The sky coordinates are corrected for the small offset with respect to GAIA-corrected literature positions of the central quasars for legacy and better comparison with JWST in the future. 
\begin{landscape}
\begin{table}
\caption{156 LAEs detected in this work with their respective quasar field from which they are detected, IDs assigned to them, sky coordinates (corrected for astrometry based on GAIA corrected literature positions for the quasars), Lyman $\alpha$ emission redshift ($z_\text
{Lya}$), systemic redshift ($z_\text{systemic}$) estimated from \citet{Verhamme2018}, Ly$\alpha$ flux, doublet flag and cross-match detections in literature \citep{diaz2021} and \citep{Kashino2025}. }
\label{tab:LAE detections}
\begin{tabular}{cccccccp{2cm}p{2cm}cp{3cm}}

\hline
QSO & ID & RA & Dec & $z_\text{Lya}$ & $z_\text{systemic}$ & FWHM (\AA) & \multicolumn{1}{p{2cm}}{\centering Flux$\times10^{-20}$\\$(\text{erg}\text{s}^{-1}\text{cm}^{-2})$} & \multicolumn{1}{p{2cm}}{\centering $\sigma$Flux$\times10^{-20}$\\$(\text{erg}\text{s}^{-1}\text{cm}^{-2})$} & DOUBLET & Notes \\
\hline

        J1030+0524 & 5 & 157.6127033 & 5.421031894 & 2.933 & 2.928 & 7.09 & 476 & 136 & N & ~ \\ 
        J1030+0524 & 7 & 157.6138504 & 5.406768641 & 2.982 & 2.979 & 4.35 & 750 & 103 & N & ~ \\ 
        J1030+0524 & 8 & 157.6143503 & 5.410317133 & 2.982 & 2.978 & 5.85 & 2178 & 391 & N & ~ \\ 
        J1030+0524 & 11 & 157.6066344 & 5.415457563 & 3.022 & 3.018 & 6.09 & 397 & 150 & N & ~ \\ 
        J1030+0524 & 12 & 157.613567 & 5.418360607 & 3.053 & 3.05 & 4.83 & 1082 & 170 & N & ~ \\ 
        J1030+0524 & 13 & 157.6059215 & 5.417340242 & 3.051 & 3.05 & 3.47 & 775 & 143 & Y & ~ \\ 
        J1030+0524 & 15 & 157.6166936 & 5.421383052 & 3.087 & 3.084 & 4.02 & 423 & 99 & N & ~ \\ 
        J1030+0524 & 17 & 157.6194538 & 5.413471934 & 3.15 & 3.149 & 3.44 & 1314 & 139 & Y & ~ \\ 
        J1030+0524 & 22 & 157.603388 & 5.413020332 & 3.233 & 3.233 & 4.03 & 1845 & 237 & Y & ~ \\ 
        J1030+0524 & 23 & 157.6147417 & 5.409903662 & 3.25 & 3.249 & 4.35 & 1136 & 123 & Y & ~ \\ 
        J1030+0524 & 24 & 157.6221003 & 5.417912586 & 3.249 & 3.247 & 3.34 & 781 & 71 & N & ~ \\ 
        J1030+0524 & 27 & 157.6127466 & 5.42063348 & 3.347 & 3.341 & 8.63 & 671 & 165 & N & ~ \\ 
        J1030+0524 & 29 & 157.6182424 & 5.415217401 & 3.366 & 3.365 & 2.07 & 441 & 78 & Y & ~ \\ 
        J1030+0524 & 32 & 157.6217877 & 5.418087077 & 3.394 & 3.391 & 4.03 & 1371 & 113 & N & ~ \\ 
        J1030+0524 & 33 & 157.6206331 & 5.418634792 & 3.393 & 3.392 & 1.68 & 296 & 51 & N & ~ \\ 
        J1030+0524 & 34 & 157.6113749 & 5.420954438 & 3.423 & 3.419 & 5.2 & 437 & 89 & N & ~ \\ 
        J1030+0524 & 35 & 157.6135381 & 5.421432224 & 3.443 & 3.44 & 4.75 & 777 & 115 & N & ~ \\ 
        J1030+0524 & 36 & 157.6208166 & 5.417600159 & 3.447 & 3.446 & 2 & 168 & 57 & N & ~ \\ 
        J1030+0524 & 37 & 157.6069794 & 5.418698688 & 3.451 & 3.446 & 6.62 & 1303 & 202 & N & ~ \\ 
        J1030+0524 & 39 & 157.613918 & 5.417407024 & 3.461 & 3.459 & 2.97 & 268 & 67 & N & ~ \\ 
        J1030+0524 & 47 & 157.6074654 & 5.412887872 & 3.683 & 3.678 & 6.47 & 275 & 93 & N & ~ \\ 
        J1030+0524 & 48 & 157.6083545 & 5.417004313 & 3.699 & 3.697 & 3.77 & 572 & 117 & N & ~ \\ 
        J1030+0524 & 49 & 157.6143686 & 5.419016229 & 3.714 & 3.713 & 4.6 & 1349 & 140 & Y & ~ \\ 
        J1030+0524 & 50 & 157.6125118 & 5.420939468 & 3.717 & 3.713 & 5.65 & 717 & 137 & N & ~ \\ 
        J1030+0524 & 51 & 157.6149747 & 5.417416855 & 3.718 & 3.716 & 3.65 & 610 & 83 & N & ~ \\ 
        J1030+0524 & 53 & 157.6195527 & 5.414784274 & 3.757 & 3.755 & 2.81 & 456 & 147 & Y & ~ \\ 
        J1030+0524 & 54 & 157.6095079 & 5.413380838 & 3.786 & 3.782 & 5.44 & 1342 & 145 & N & ~ \\ 
        J1030+0524 & 57 & 157.6089956 & 5.416709289 & 3.824 & 3.823 & 6.92 & 3320 & 298 & Y & ~ \\ 
        J1030+0524 & 58 & 157.612827 & 5.40747609 & 3.823 & 3.823 & 4.12 & 1233 & 149 & Y & ~ \\ 
        J1030+0524 & 63 & 157.6151402 & 5.40802336 & 3.962 & 3.962 & 6.34 & 7026 & 202 & Y & ~ \\ 
        J1030+0524 & 64 & 157.6131062 & 5.420431599 & 3.964 & 3.961 & 4.71 & 1022 & 106 & N & ~ \\ 
        J1030+0524 & 67 & 157.6162873 & 5.411687751 & 4.117 & 4.117 & 2.95 & 501 & 116 & Y & ~ \\ 
        J1030+0524 & 69 & 157.6064293 & 5.413087752 & 4.161 & 4.158 & 4.74 & 280 & 136 & N & ~ \\ 
        J1030+0524 & 70 & 157.6124544 & 5.412706364 & 4.201 & 4.2 & 3.55 & 1010 & 93 & Y & ~ \\ 
        J1030+0524 & 87 & 157.6104259 & 5.424664863 & 4.439 & 4.438 & 2.19 & 634 & 59 & N & ~ \\ 
        J1030+0524 & 88 & 157.6159553 & 5.406162462 & 4.53 & 4.527 & 4.99 & 1685 & 84 & N & ~ \\ 
        J1030+0524 & 90 & 157.6084311 & 5.419252729 & 4.547 & 4.546 & 3.34 & 409 & 44 & N & ~ \\ 
        J1030+0524 & 99 & 157.6215307 & 5.417304454 & 4.708 & 4.706 & 3.61 & 567 & 67 & N & ~ \\ 
        J1030+0524 & 100 & 157.6190324 & 5.415102373 & 4.709 & 4.707 & 3.13 & 380 & 62 & N & ~ \\ 
        J1030+0524 & 101 & 157.6173202 & 5.412332886 & 4.938 & 4.937 & 2.99 & 172 & 50 & N & ~ \\ 
        J1030+0524 & 102 & 157.6189916 & 5.419800153 & 4.952 & 4.95 & 3.41 & 169 & 150 & N & D21 LAE\#1 \\

\end{tabular}   
\end{table}
\end{landscape}

\begin{landscape}
    \begin{table}
        \contcaption{}
        \begin{tabular}{cccccccp{2cm}p{2cm}cp{3cm}}
        \hline
QSO & ID & RA & Dec & $z_\text{Lya}$ & $z_\text{systemic}$  & FWHM (\AA) & \multicolumn{1}{p{2cm}}{\centering Flux$\times10^{-20}$\\$(\text{erg}\text{s}^{-1}\text{cm}^{-2})$} & \multicolumn{1}{p{2cm}}{\centering $\sigma$Flux$\times10^{-20}$\\$(\text{erg}\text{s}^{-1}\text{cm}^{-2})$} & DOUBLET & Notes \\
\hline
        
         J1030+0524 & 103 & 157.615194 & 5.420705944 & 4.951 & 4.95 & 2.39 & 211 & 38 & N & ~ \\ 
        J1030+0524 & 107 & 157.6093443 & 5.422806641 & 5.109 & 5.105 & 6.29 & 570 & 113 & N & ~ \\ 
        J1030+0524 & 109 & 157.6081566 & 5.417402659 & 5.152 & 5.145 & 10.05 & 700 & 114 & N & ~ \\ 
        J1030+0524 & 110 & 157.6086986 & 5.410447489 & 5.163 & 5.158 & 7.95 & 547 & 104 & N & ~ \\ 
        J1030+0524 & 111 & 157.6155935 & 5.414579005 & 5.195 & 5.194 & 2.88 & 539 & 97 & Y & ~ \\ 
        J1030+0524 & 113 & 157.6055184 & 5.414038933 & 5.225 & 5.218 & 9.15 & 2163 & 92 & N & ~ \\ 
        J1030+0524 & 118 & 157.6210706 & 5.41879523 & 5.277 & 5.269 & 10.55 & 3162 & 143 & N & ~ \\ 
        J1030+0524 & 119 & 157.6123039 & 5.407709629 & 5.341 & 5.339 & 4.32 & 301 & 74 & N & ~ \\ 
        J1030+0524 & 120 & 157.6196941 & 5.413626843 & 5.375 & 5.374 & 4.37 & 1761 & 267 & Y & ~ \\ 
        J1030+0524 & 122 & 157.6129968 & 5.422094561 & 5.433 & 5.428 & 7.82 & 752 & 110 & N & ~ \\ 
        J1030+0524 & 123 & 157.6078562 & 5.412576469 & 5.449 & 5.448 & 2.18 & 213 & 81 & N & ~ \\ 
        J1030+0524 & 124 & 157.6151371 & 5.419909315 & 5.455 & 5.454 & 1.96 & 366 & 35 & N & ~ \\ 
        J1030+0524 & 125 & 157.61107 & 5.415640883 & 5.521 & 5.514 & 9.92 & 1906 & 267 & N & D21 LAE\#3, K26 OIII\#7771 \\ 
        J1030+0524 & 126 & 157.609656 & 5.419318972 & 5.521 & 5.519 & 4.15 & 545 & 77 & N & D21 LAE\#2, K26 OIII\#8631 \\ 
        J1030+0524 & 127 & 157.6089866 & 5.411184503 & 5.53 & 5.529 & 2.88 & 249 & 53 & N & ~ \\ 
        J1030+0524 & 128 & 157.6105245 & 5.409785174 & 5.528 & 5.526 & 3.94 & 228 & 121 & N & D21 LAE\#4 \\ 
        J1030+0524 & 129 & 157.611719 & 5.413976496 & 5.546 & 5.543 & 5.05 & 280 & 56 & N & ~ \\ 
        J1030+0524 & 131 & 157.6152974 & 5.405450429 & 5.727 & 5.721 & 8.91 & 1606 & 120 & N & D21 LAE\#5, K26 OIII\#6714 \\ 
        J1030+0524 & 132 & 157.6124061 & 5.4155049 & 5.728 & 5.724 & 6 & 246 & 80 & N & D21 LAE\#6, K26 OIII\#7377 \\ 
        J1030+0524 & 134 & 157.6136202 & 5.414949557 & 5.747 & 5.741 & 8.92 & 452 & 122 & N & K26 OIII\#6949 \\ 
        J1030+0524 & 137 & 157.6206675 & 5.414715582 & 5.762 & 5.756 & 8.85 & 945 & 116 & N & D21 LAE\#7 \\ 
        J1030+0524 & 138 & 157.6131874 & 5.418853694 & 5.763 & 5.758 & 8.06 & 591 & 117 & N & D21 LAE\#8, K26 LAE\#7493 \\ 
        J1030+0524 & 139 & 157.6168812 & 5.415048523 & 5.762 & 5.759 & 4.68 & 289 & 112 & N & ~ \\ 
        J1030+0524 & 140 & 157.6102838 & 5.414320258 & 5.766 & 5.756 & 14.68 & 730 & 165 & N & K26 OIII\#7799 \\ 
        J1030+0524 & 141 & 157.6087895 & 5.413954246 & 5.772 & 5.76 & 17.52 & 878 & 173 & N & K26 OIII\#8308 \\ 
        J1030+0524 & 157 & 157.611157 & 5.419536205 & 6.61 & 6.603 & 11 & 2082 & 278 & N & ~ \\ 
        J1030+0524 & 158 & 157.6166214 & 5.412154112 & 6.022 & 6.021 & 3.46 & 437 & 103 & N & ~ \\ 
        J1030+0524 & 172 & 157.6145606 & 5.40690073 & 6.655 & 6.651 & 6.86 & 2902 & 1589 & N & ~ \\ 
        J1030+0524 & 173 & 157.615903 & 5.410462657 & 6.675 & 6.675 & 2.19 & 2393 & 248 & N & ~ \\ 
        J1030+0524 & 174 & 157.6079773 & 5.412041752 & 6.675 & 6.672 & 5 & 2322 & 1770 & N & ~ \\ 
        J1306+0356 & 6 & 196.5356347 & 3.946543734 & 2.983 & 2.972 & 14.26 & 982 & 364 & N & ~ \\ 
        J1306+0356 & 8 & 196.5311851 & 3.944712887 & 2.994 & 2.991 & 4.7 & 2360 & 273 & N & ~ \\ 
        J1306+0356 & 9 & 196.5328446 & 3.944279734 & 2.997 & 2.993 & 5.78 & 4332 & 392 & N & ~ \\ 
        J1306+0356 & 10 & 196.5392224 & 3.934305021 & 2.998 & 2.993 & 7.03 & 1748 & 180 & N & ~ \\ 
        J1306+0356 & 11 & 196.5411429 & 3.942276704 & 2.997 & 2.993 & 5.7 & 1030 & 199 & N & ~ \\ 
        J1306+0356 & 12 & 196.5330608 & 3.943545102 & 2.999 & 2.995 & 6.1 & 913 & 125 & N & ~ \\ 
        J1306+0356 & 15 & 196.5407919 & 3.945295868 & 3.024 & 3.023 & 1.88 & 400 & 74 & N & ~ \\ 
        J1306+0356 & 16 & 196.5399075 & 3.938491335 & 3.065 & 3.06 & 7.54 & 675 & 287 & N & ~ \\ 
        J1306+0356 & 19 & 196.5398721 & 3.943108247 & 3.175 & 3.175 & 3.88 & 1967 & 110 & Y & ~ \\ 
        J1306+0356 & 20 & 196.5351301 & 3.933834901 & 3.177 & 3.172 & 7.88 & 1302 & 206 & N & ~ \\  
        
        \end{tabular}
        
        \label{tab:LAE detctions2}
    \end{table}
\end{landscape}

\begin{landscape}
    \begin{table}
    \contcaption{}
        \begin{tabular}{cccccccp{2cm}p{2cm}cp{3cm}}
        \hline
QSO & ID & RA & Dec & $z_\text{Lya}$ & $z_\text{systemic}$  & FWHM (\AA) & \multicolumn{1}{p{2cm}}{\centering Flux$\times10^{-20}$\\$(\text{erg}\text{s}^{-1}\text{cm}^{-2})$} & \multicolumn{1}{p{2cm}}{\centering $\sigma$Flux$\times10^{-20}$\\$(\text{erg}\text{s}^{-1}\text{cm}^{-2})$} & DOUBLET & Notes \\
\hline
         J1306+0356 & 21 & 196.5297357 & 3.944363679 & 3.175 & 3.174 & 5.41 & 3090 & 310 & Y & ~ \\ 
        J1306+0356 & 22 & 196.5400453 & 3.947718418 & 3.199 & 3.195 & 7 & 6074 & 267 & N & ~ \\ 
        J1306+0356 & 23 & 196.5402755 & 3.943280705 & 3.201 & 3.196 & 7.51 & 6549 & 551 & N & ~ \\ 
        J1306+0356 & 24 & 196.531506 & 3.934170464 & 3.224 & 3.221 & 4.28 & 1292 & 178 & N & ~ \\ 
        J1306+0356 & 26 & 196.5357978 & 3.93715352 & 3.319 & 3.316 & 4.24 & 6079 & 205 & N & ~ \\ 
        J1306+0356 & 29 & 196.5339331 & 3.945161899 & 3.472 & 3.468 & 4.86 & 371 & 101 & N & ~ \\ 
        J1306+0356 & 30 & 196.5362777 & 3.934906777 & 3.479 & 3.477 & 3.33 & 893 & 144 & N & ~ \\ 
        J1306+0356 & 33 & 196.5411537 & 3.944378902 & 3.544 & 3.539 & 8.44 & 2276 & 280 & N & ~ \\ 
        J1306+0356 & 34 & 196.5392266 & 3.946822823 & 3.556 & 3.552 & 6.92 & 6040 & 640 & N & ~ \\ 
        J1306+0356 & 37 & 196.5342686 & 3.946449138 & 3.709 & 3.707 & 3.81 & 432 & 120 & N & ~ \\ 
        J1306+0356 & 38 & 196.5314012 & 3.93487056 & 4.043 & 4.039 & 5.5 & 666 & 123 & N & ~ \\ 
        J1306+0356 & 45 & 196.5305648 & 3.941654644 & 3.917 & 3.916 & 4 & 573 & 181 & Y & ~ \\ 
        J1306+0356 & 48 & 196.527972 & 3.935219164 & 4.045 & 4.033 & 17.9 & 609 & 140 & N & ~ \\ 
        J1306+0356 & 54 & 196.5405706 & 3.945585504 & 4.105 & 4.096 & 12.02 & 383 & 128 & N & ~ \\ 
        J1306+0356 & 69 & 196.5302834 & 3.938452354 & 4.426 & 4.425 & 1.98 & 952 & 86 & N & ~ \\ 
        J1306+0356 & 70 & 196.537045 & 3.9339601 & 4.431 & 4.429 & 3.58 & 332 & 64 & N & ~ \\ 
        J1306+0356 & 72 & 196.5319775 & 3.939789972 & 4.472 & 4.471 & 5.79 & 1472 & 172 & Y & ~ \\ 
        J1306+0356 & 73 & 196.5318952 & 3.939484101 & 4.471 & 4.465 & 8.77 & 1285 & 201 & N & ~ \\ 
        J1306+0356 & 74 & 196.5415734 & 3.933757055 & 4.52 & 4.517 & 4.54 & 162 & 144 & N & ~ \\ 
        J1306+0356 & 78 & 196.528194 & 3.940935019 & 4.616 & 4.61 & 9.44 & 924 & 275 & N & ~ \\ 
        J1306+0356 & 88 & 196.5388822 & 3.943888751 & 4.664 & 4.658 & 10.11 & 489 & 112 & N & ~ \\ 
        J1306+0356 & 90 & 196.5277855 & 3.934668252 & 4.711 & 4.711 & 1.79 & 235 & 46 & N & ~ \\ 
        J1306+0356 & 98 & 196.5306933 & 3.943387483 & 4.862 & 4.854 & 11.91 & 476 & 151 & N & ~ \\ 
        J1306+0356 & 102 & 196.5392379 & 3.935835113 & 4.935 & 4.932 & 4.61 & 161 & 59 & N & ~ \\ 
        J1306+0356 & 122 & 196.5346943 & 3.934446653 & 5.111 & 5.105 & 8.67 & 2310 & 148 & N & ~ \\ 
        J1306+0356 & 124 & 196.5324361 & 3.937887853 & 5.114 & 5.113 & 2.55 & 550 & 76 & Y & ~ \\ 
        J1306+0356 & 128 & 196.5316355 & 3.940925094 & 5.175 & 5.173 & 2.85 & 179 & 60 & N & ~ \\ 
        J1306+0356 & 134 & 196.5356285 & 3.940669312 & 5.221 & 5.211 & 14.31 & 668 & 121 & N & ~ \\ 
        J1306+0356 & 142 & 196.5316602 & 3.936749559 & 5.343 & 5.342 & 7.38 & 2778 & 165 & Y & ~ \\ 
        J1306+0356 & 151 & 196.5282358 & 3.941403978 & 5.358 & 5.353 & 8.57 & 1493 & 118 & N & ~ \\ 
        J1306+0356 & 152 & 196.5328849 & 3.945078924 & 5.357 & 5.357 & 4.68 & 850 & 116 & Y & ~ \\ 
        J1306+0356 & 154 & 196.5355984 & 3.947652127 & 5.431 & 5.427 & 5.93 & 512 & 128 & N & ~ \\ 
        J1306+0356 & 187 & 196.5302233 & 3.937163585 & 5.632 & 5.632 & 2.26 & 301 & 78 & N & ~ \\ 
        J1306+0356 & 191 & 196.5387515 & 3.945981944 & 5.725 & 5.716 & 13.45 & 595 & 150 & N & ~ \\ 
        J1306+0356 & 192 & 196.5321757 & 3.941165883 & 5.764 & 5.756 & 11.77 & 356 & 135 & N & ~ \\ 
        J1306+0356 & 207 & 196.5334964 & 3.94019048 & 5.972 & 5.97 & 2.87 & 425 & 44 & N & ~ \\ 
        J1306+0356 & 211 & 196.5331677 & 3.941036253 & 6.033 & 6.017 & 23.24 & 2697 & 395 & N & ~ \\ 
        J1306+0356 & 212 & 196.5408454 & 3.940484045 & 6.038 & 6.029 & 12.04 & 851 & 270 & N & ~ \\ 
        J1306+0356 & 214 & 196.5347462 & 3.938884384 & 6.079 & 6.077 & 3.81 & 548 & 54 & N & ~ \\ 
        J1306+0356 & 217 & 196.535704 & 3.944538755 & 6.175 & 6.167 & 12.09 & 1977 & 94 & N & ~ \\ 
        J1306+0356 & 249 & 196.5383533 & 3.93707687 & 6.418 & 6.411 & 10.66 & 467 & 214 & N & ~ \\ 
        J1306+0356 & 250 & 196.535274 & 3.937247278 & 6.42 & 6.414 & 9.19 & 742 & 457 & N & ~ \\ 
        J1306+0356 & 254 & 196.5375781 & 3.944206544 & 6.48 & 6.479 & 2.52 & 375 & 62 & N & ~ \\ 
        J1306+0356 & 257 & 196.5385563 & 3.946970563 & 6.517 & 6.517 & 2.1 & 252 & 53 & N & ~ \\

        \end{tabular}
        
        \label{tab:LAE detections3}
    \end{table}
\end{landscape}

\begin{landscape}
    \begin{table}
        \contcaption{}
        \begin{tabular}{cccccccp{2cm}p{2cm}cp{3cm}}
        \hline
QSO & ID & RA & Dec & $z_\text{Lya}$ & $z_\text{systemic}$  & FWHM (\AA) & \multicolumn{1}{p{2cm}}{\centering Flux$\times10^{-20}$\\$(\text{erg}\text{s}^{-1}\text{cm}^{-2})$} & \multicolumn{1}{p{2cm}}{\centering $\sigma$Flux$\times10^{-20}$\\$(\text{erg}\text{s}^{-1}\text{cm}^{-2})$} & DOUBLET & Notes \\
\hline
         J1306+0356 & 260 & 196.5333827 & 3.940873938 & 6.518 & 6.516 & 4.96 & 251 & 114 & N & ~ \\ 
        J1306+0356 & 264 & 196.5369489 & 3.94087956 & 6.552 & 6.551 & 2.84 & 270 & 58 & N & ~ \\ 
        J1306+0356 & 265 & 196.5335637 & 3.94143116 & 6.573 & 6.572 & 2.45 & 212 & 144 & N & ~ \\ 
        J1306+0356 & 266 & 196.5398205 & 3.937422327 & 6.585 & 6.584 & 3.32 & 506 & 110 & Y & ~ \\ 
        J1306+0356 & 268 & 196.5392656 & 3.940855947 & 6.64 & 6.64 & 8.99 & 982 & 278 & Y & ~ \\ 
        J158-14 & 3 & 158.7015353 & -14.41698612 & 3.072 & 3.068 & 5.71 & 2251 & 388 & N & ~ \\ 
        J158-14 & 7 & 158.6944473 & -14.42127994 & 3.24 & 3.236 & 5.71 & 1212 & 273 & N & ~ \\ 
        J158-14 & 10 & 158.6943713 & -14.41390877 & 3.586 & 3.585 & 1.49 & 2937 & 330 & N & ~ \\ 
        J158-14 & 11 & 158.6938701 & -14.42716519 & 3.586 & 3.584 & 2.85 & 2458 & 377 & N & ~ \\ 
        J158-14 & 15 & 158.6966724 & -14.41449514 & 4.003 & 3.999 & 6.08 & 880 & 275 & N & ~ \\ 
        J158-14 & 16 & 158.7007908 & -14.41639699 & 4.022 & 4.022 & 3.75 & 1963 & 258 & Y & ~ \\ 
        J158-14 & 21 & 158.7004878 & -14.42264593 & 4.497 & 4.497 & 5.79 & 1043 & 301 & Y & ~ \\ 
        J158-14 & 24 & 158.7013235 & -14.41456814 & 4.551 & 4.548 & 6.15 & 1158 & 168 & N & ~ \\ 
        J158-14 & 41 & 158.7000772 & -14.41950856 & 5.104 & 5.1 & 6.42 & 715 & 178 & N & ~ \\ 
        J158-14 & 42 & 158.6893233 & -14.42804053 & 5.125 & 5.124 & 3.05 & 1061 & 193 & Y & ~ \\ 
        J158-14 & 43 & 158.6919866 & -14.42578878 & 5.219 & 5.217 & 4.03 & 1095 & 183 & N & ~ \\ 
        J158-14 & 51 & 158.6931437 & -14.42633309 & 5.259 & 5.259 & 11.1 & 3506 & 563 & N & ~ \\ 
        J158-14 & 53 & 158.6955444 & -14.4227409 & 5.302 & 5.301 & 2.83 & 1069 & 234 & Y & ~ \\ 
        J158-14 & 55 & 158.6902089 & -14.42620183 & 5.357 & 5.355 & 3.98 & 428 & 127 & N & ~ \\ 
        J158-14 & 61 & 158.6917644 & -14.42196371 & 5.481 & 5.477 & 5.92 & 743 & 143 & N & ~ \\ 
        J158-14 & 67 & 158.6995279 & -14.41511168 & 5.754 & 5.747 & 10.7 & 2344 & 155 & N & ~ \\ 
        J158-14 & 85 & 158.6982974 & -14.4220621 & 6.308 & 6.307 & 2.15 & 504 & 107 & N & ~ \\ 
        J158-14 & 108 & 158.6891468 & -14.4163953 & 6.406 & 6.405 & 3.53 & 2864 & 194 & Y & ~ \\ 
        J158-14 & 114 & 158.6883109 & -14.41740684 & 6.498 & 6.491 & 11.15 & 1363 & 276 & N & ~ \\ 
        J158-14 & 120 & 158.6935183 & -14.41880374 & 6.517 & 6.513 & 6.17 & 1386 & 323 & N & ~ \\ 
        J158-14 & 128 & 158.6932626 & -14.42026465 & 6.543 & 6.526 & 24.71 & 1471 & 227 & N & ~ \\ 
        J158-14 & 130 & 158.6933468 & -14.42160668 & 6.55 & 6.546 & 7.6 & 442 & 114 & N & ~ \\ 
        J158-14 & 131 & 158.689095 & -14.42230834 & 6.553 & 6.552 & 2.52 & 225 & 141 & N & ~ \\ 
        J158-14 & 132 & 158.6934617 & -14.42231465 & 6.549 & 6.547 & 4.02 & 453 & 419 & N & ~ \\ 
        J158-14 & 134 & 158.6978597 & -14.42375167 & 6.554 & 6.552 & 4 & 228 & 127 & N & ~ \\ 
        J158-14 & 136 & 158.6923287 & -14.41453629 & 6.582 & 6.58 & 3.93 & 1288 & 259 & N & ~ \\ \hline
        \end{tabular}
        
        \label{tab:LAE detections4}
    \end{table}
\end{landscape}

\section{Completeness of LAE detection}
\label{sec:completeness}
The completeness of the detected sample of LAEs is a crucial metric to understand the fraction of real LAEs that are recovered from the MUSE cubes using the above procedures. It measures the success of the software process in detecting, identifying and fitting the emission lines. 

In order to assess the completeness of the detected LAE sample, mock cubes were created with no continuum or detected sources. The mock cubes reflect the noise properties of the original cubes and then fake sources are inserted at random x and y positions in the cube. The redshifts at which these sources are placed are also drawn randomly from a range of redshifts between the minimum and maximum redshifts of the actual LAE detections. The x,y and z pixels that have been masked while processing the original cube are avoided while inserting the fake sources. Similarly, random values of velocity dispersion ranging from 100-400 km/s are assigned to each fake source. The fluxes of the fake sources range from log $F[\text{erg s}^{-1}\text{cm}^{-2}]=-18$ to $F=-14.9$. 100 fake sources with flux values spaced at $\sim0.03$ dex and randomly assigned redshift, velocity dispersion and sky coordinates are inserted into each mock cube. 

Each fake LAEs can be created using a 1D Gaussian emission line. 
The 1D Gaussian is a skewed one (see equation \ref{eq:gaussian}) where the asymmetry of the profile is kept fixed for all fake sources. A skewness value of 2.48 is used which is the mean skewness for the LAEs detected in this work. 

The flux at each wavelength is distributed spatially over a 2D Gaussian. For the standard deviation, we adopt the mean isophotal radius measured for all LAEs in LSDCat.
The isophotal radius will encompass 68\% of the source flux for a point source blurred by a Gaussian point spread function (PSF) \citep{LSDCat1.0}. 

\begin{figure}
    \centering
    \includegraphics[width=\linewidth]{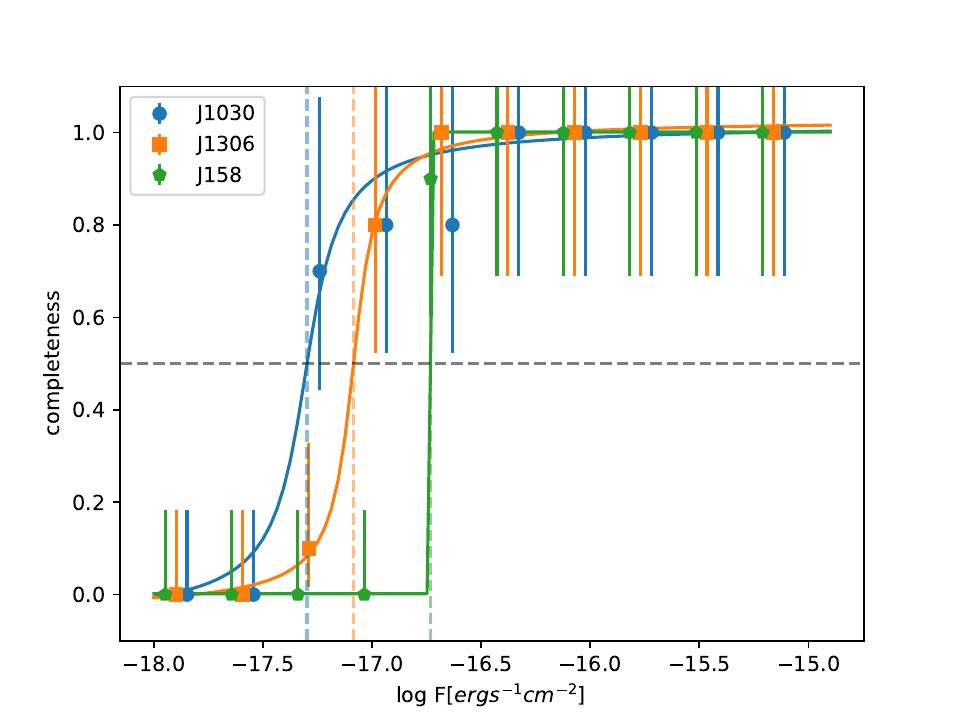}
    \caption{The completeness of LAE sample as a function of log $F$. The data points from each cube are offset for visual clarity. The completeness curve is estimated for each cube separately and is colour coded accordingly. The 50\% completeness limit is also shown using dotted vertical lines.}
    \label{fig:completeness}
\end{figure}

The mock cubes are then treated exactly the same way as the actual MUSE datacubes by passing them through LSDCat and QtClassify to understand how many of the inserted fake sources are recovered after detection and classification of the emission lines. The completeness of our LAE sample as a function of log $F$ is shown in Figure \ref{fig:completeness}. Plotted here are the mid-points of the flux bins, encompassing fluxes in the range log $F[\text{erg s}^{-1}\text{cm}^{-2}]=-18$ to -14.9. The 1$\sigma$ Poisson errors on these completeness estimates are calculated using formulae from \citet{Gehrels1986CONFIDENCEDATA}. Since each of the cubes studied in this work has different noise properties, their completeness is assessed separately. J1030 (blue) has the highest completeness while J158 (green), owing to its higher noise level, has the lowest cut-off owing to the lower depth of the data. This is in agreement with the observed root mean square value of the cube (Meyer et al. in prep) which is significantly higher than J1030 and J1306. The dashed vertical lines mark the 50\% completeness limit for each cube and the values can be seen in Table \ref{tab:arctan fit}. Each completeness curve is fit using an arctan function given by
\begin{equation}
\label{eq:arctan}
    \text{Completeness(log }F) = S_y\text{arctan}(S_x(\text{log }F-T_x))+T_y
\end{equation} where $T_x$ and $T_y$ defines the translation of the curve along each axes while $S_x$ and $S_y$ determines the scaling along each direction. The best fit parameters for each datacube are given in Table \ref{tab:arctan fit}.

\begin{table*}
    \centering
    \caption{The best fit parameters for the arctan function (equation \ref{eq:arctan}) used to fit the completeness curves. The 50\% completeness limits for each cube are also given here.}
    \begin{tabular}{c|c|c|c|c|p{3cm}|}
    \hline
        QSO & $S_y$ & $S_x$ & $T_x$ & $T_y$ & \multicolumn{1}{|p{3cm}|}{\centering 50\% completeness limit\\log $F[\text{ergs}^{-1}\text{cm}^{-2}]$}\\
        \hline
        SDSSJ1030+0524 & 0.34 & 8.78 & -17.3 & 0.48 & \multicolumn{1}{|p{3cm}|}{\centering -17.29}\\
        SDSSJ1306+0356 & -0.34 & -13.0 & -17.0 & 0.49 & \multicolumn{1}{|p{3cm}|}{\centering -17.03}\\
        PSOJ158-14 & -0.32 & $-6.83\times10^5$ & -16.6 & 0.50 & \multicolumn{1}{|p{3cm}|}{\centering -16.63}\\
        \hline
    \end{tabular}
    
    \label{tab:arctan fit}
\end{table*}

\section{LAE Luminosity Function at \texorpdfstring{$2.9<z<6.7$}{3lezle6}}
\label{subsec:luminosity}
We detected 156 LAEs detected in three $z\sim6$ quasar fields with luminosity ranging from log $L[\text{ergs}^{-1}]=41.3$ to $44$ at $2.9<z<6.7$. To estimate the luminosity function (LF) of the LAEs detected in the MUSE datacubes, we adopt the $1/V$ estimator method used in \citet{drakehdf2017, drake2017}. The $1/V$ estimator makes no prior assumption on the shape of the function which gives the number density of objects in bins of luminosity. For an LAE, i, the maximum comoving volume within which it can be observed is given by 
\begin{equation}
\label{{eq:V_max}}
    V_{\text{max}}(L_i,z) = \int_{z1}^{z2}\frac{dV}{dz}C(L_i,z)dz
\end{equation}where $z_1$ and $z_2$ are minimum and maximum redshifts of the luminosity bin, d$V$ is the comoving volume element corresponding to redshift interval d$z=0.01$ and $C(L,z)$ is the completeness curve for an object with luminosity $L_i$ across all redshifts $z_i$. The completeness corresponding to the LAE with luminosity $L_i$ at $2.9<z<6.7$ can be obtained using equation \ref{eq:arctan} corresponding to the quasar field where the LAE is detected. We probe  a total volume of $1.98\times10^4~\text{Mpc}^3$ using the three quasar fields from MUSE. When calculating LF, the masked regions are completely excluded from all parts of the analysis including when calculating the volume, and thus these regions are also excluded from the completeness calculations.

\begin{figure}
     \centering
     \includegraphics[width=\linewidth]{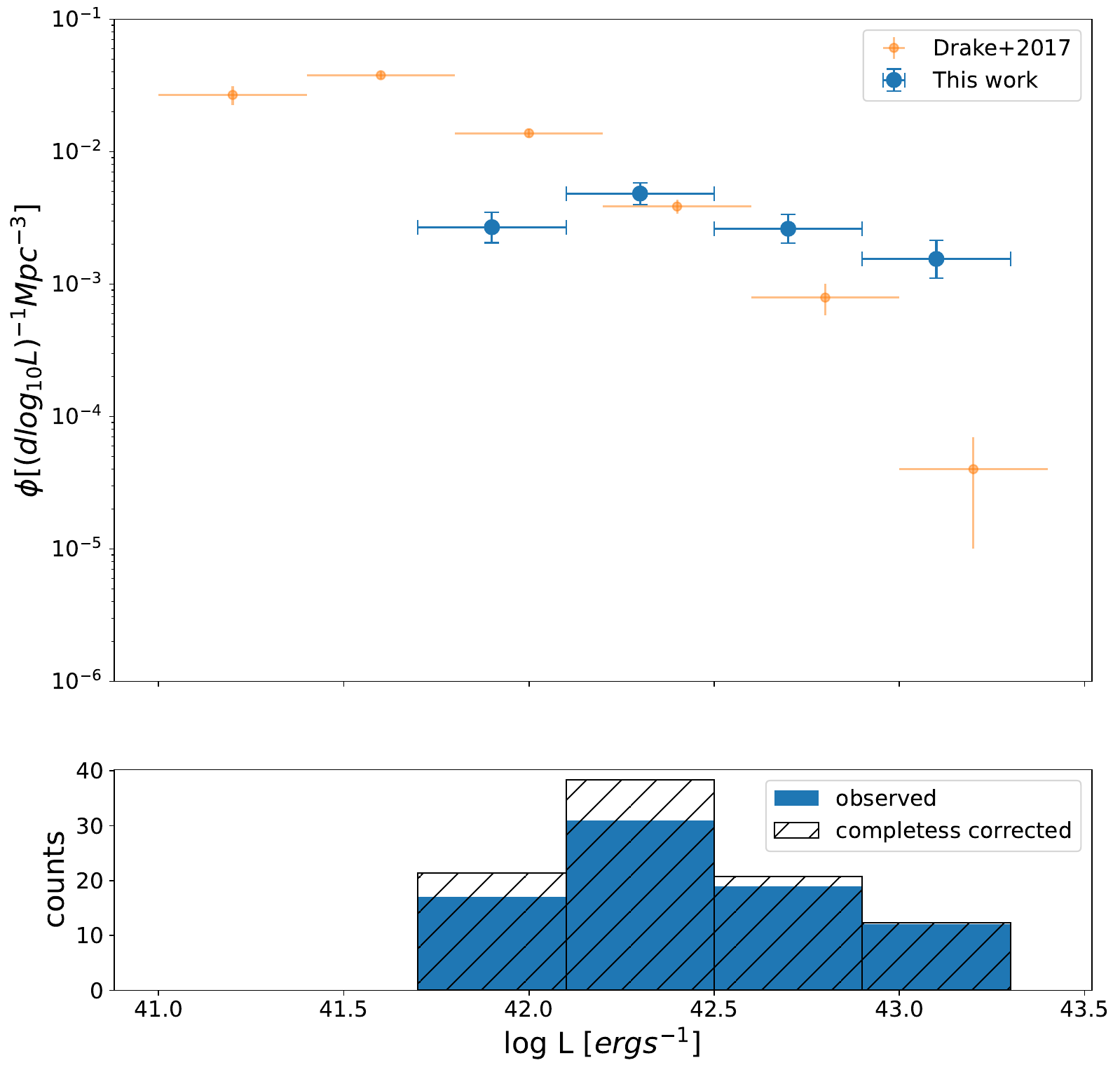}
     \caption{\textit{Upper panel}: LF for 71 LAEs ($>50\%$ complete) detected at $2.9<z<6.7$ using three MUSE fields. The results obtained in this work are compared to the luminosity function from \citet{drake2017} denoted using orange points. A Schechter function is used to fit the data shown using the blue curve. \textit{Lower panel}: Histogram showing the number of LAEs (both observed and completeness corrected) in each log $L$ bin.}
     \label{fig:LF}
 \end{figure}

 The number density of objects per bin is then computed using
 \begin{equation}
 \label{eq:luminosity}
     \phi[(\text{dlog}_{10}L)^{-1}\text{Mpc}^{-3}]=\sum_i\frac{1}{V_{\text{max}}(L_i,z)}/\text{bin size}
 \end{equation} by substituting for $V_\text{max}$ calculated for each LAE. The log $L$ bin size used here is 0.3 dex. Given that the completeness function for J158 is low, we have only included LAEs with $>50\%$ completeness in all cubes (see Figure \ref{fig:completeness}) to avoid overestimating the number density of LAEs in each luminosity bin. There are 73 out of 156 LAEs that are above the 50\% completeness limit with luminosities of $41.7<\text{log}~L[\text{ergs}^{-1}]<43.2$. The LF values obtained for these LAEs detected in three quasar fields is given in Table \ref{tab:LF}
 \begin{table*}
     \centering
     \caption{LAE LF values obtained for the six log $L$ bins used in this work using $1/V_\text{max}$ method. The median values of each log $L$ bin, the number of LAEs detected in each bin without completeness correction and the 1$\sigma$ error corresponding to each LF value are given here. }
     \begin{tabular}{c|c|c|c|c}
     \hline
      log $L~[\text{ergs}^{-1}]$ bins & mid log $L~[\text{ergs}^{-1}]$  & No. of LAEs & $\phi [(\text{dlog}_{10}L)^{-1}\text{Mpc}^{-3}]$ & $\delta\phi [(\text{dlog}_{10}L)^{-1}\text{Mpc}^{-3}]$(+,-)\\
      \hline
       41.7-42.1   & 41.9 & 15 & 0.0023 & 0.0008, 0.0006\\
       42.1-42.5 & 42.3 & 29 & 0.0046 & 0.0009, 0.0008\\
       42.5-42.9 & 42.7 & 17 & 0.0023 & 0.0007, 0.0005\\
       42.9-43.3 & 43.1 & 12 & 0.0015 & 0.0006, 0.0004
     \end{tabular}
     
     \label{tab:LF}
 \end{table*} and plotted in the upper panel in Figure \ref{fig:LF}. The 1 $\sigma$ confidence levels for the LF are calculated using the approximations from \citet{Gehrels1986CONFIDENCEDATA}.
The blue points represent the number density of LAEs from this work for each luminosity bin calculated using equation \ref{eq:luminosity} while the orange points represent LF values from \citet{drake2017} (\citetalias{drake2017} from here onward). 
The lower panel shows the observed counts in blue colour and the hatched bins indicate the completeness corrected bins. 

\citetalias{drake2017} detected 604 LAEs in the Hubble Ultra Deep Field (HUDF) observed using MUSE for 137 hours at a similar redshift range of this study ($2.91<z<6.64)$. Owing to the extended exposure time, they successfully identified numerous LAEs, including the low surface brightness extended Lyman $\alpha$ emissions across the probed redshift ranges. This led to a greater number density of LAEs at the faint end compared to the LF presented in this paper.  Due to the 50\% completeness cut employed in our work to account for the low completeness of one of the datacubes, the obtained LF in our work do not extend to the lowest luminosity bins in \citetalias{drake2017} which would otherwise cover the whole luminosity range used in the latter. 
However, we have an excess towards the bright end of the LF compared to \citetalias{drake2017}. This work reports 24 LAEs with log $L[\text{ergs}^{-1}]>42.6$ while \citetalias{drake2017} have only 16 in this luminosity range. The redshifts of these LAEs range between $z=3.19$ and $6.67$ with a median of $z=5.36$. The significant excess of LAEs in this work with respect to \citetalias{drake2017} (the error bars do not overlap) might be due to cosmic variance.


\bsp	
\label{lastpage}
\end{document}